\documentstyle[11pt,aaspp4]{article}

\received{June 30, 1997}

\slugcomment{Submitted to The Astrophysical Journal, 1997}

\begin{document}

\def\fd{IRAS F10214}
\def\fz{IRAS 09104}
\def\fq{IRAS F15307}
\def\mmu{$m^{-1}$}
\def\mmum{$m^{-1/2}$}
\def\hmd{$h^{-2}$}
\def\hmu{$h^{-1}$}


\title{EARLY-TYPE GALAXIES IN THE HUBBLE DEEP FIELD: THE STAR FORMATION
HISTORY$^{(\dagger)}$ \footnote {($\dagger$)
\it Based on observations made at the Kitt Peak National Observatory, 
National Optical Astronomy Observatories, which is operated by the Association 
of Universities for Research in Astronomy Inc. (AURA) under cooperative 
agreement with the National Science Foundation.}
}

\author{Alberto Franceschini\altaffilmark{1}}
\affil{Dipartimento di Astronomia di Padova, Padova, Italy}

\author{Laura Silva\altaffilmark{2}}
\affil{International School for Advanced Studies}

\author{Giovanni Fasano\altaffilmark{3}, 
Gian Luigi\ Granato\altaffilmark{3} 
and Alessandro Bressan\altaffilmark{3}}
\affil{Osservatorio Astronomico di Padova, Padova, Italy}

\author{Stephane Arnouts\altaffilmark{1}}
\affil{Dipartimento di Astronomia di Padova, Padova, Italy}

\author{Luigi\ Danese\altaffilmark{2}}
\affil{International School for Advanced Studies}

\altaffiltext{1}{Dipartimento di Astronomia, Vicolo Osservatorio 5, I-35122,
Padova, Italy. E-mail: Franceschini@pd.astro.it}
\altaffiltext{2}{International School for Advanced Studies, Via Beirut, 
I34014 Trieste, Italy}
\altaffiltext{3}{Osservatorio Astronomico, Vicolo dell'Osservatorio 5,
I35122 Padova, Italy. }

\begin{abstract}

We have investigated the properties of a complete K-band selected sample of 35
elliptical and S0 galaxies brighter than $K=20^m.15$ in the Hubble Deep Field, as
representative of the field galaxy population.  This sample has been derived from
deep K-band image by the KPNO-IRIM camera, by
applying a rigorous morphological classification scheme based on quantitative
analyses of the surface brightness profiles. The completeness of the sample is
proven by a careful evaluation of all biasing effects inherent in the automated
selection procedure. Fifteen objects have spectroscopic redshifts, while for the
remaining twenty a photometric redshift is estimated from a seven-colour broad-band
spectrum (including 4 HST and 3 near-IR bands).  This dataset, based on public
archives from HST and from deep observations at Kitt-Peak and Hawaii, is unique as
for the morphological information, and for the photometric and spectroscopic
coverage. The broad-band spectra of the sample galaxies, together with a few basic
assumptions about the IMF and the stellar evolutionary paths, allow us to date their
dominant stellar populations. The majority of bright early-type galaxies in this
field are found at redshifts $z \lesssim 1.3$ to display colors indicative of a
fairly wide range of ages (typically 1.5 to 3 Gyrs). Because of the different
cosmological timescales, the star-formation history depends to some extent on the
assumed value for the cosmological deceleration parameter: we find that the major
episodes of star-formation building up typical $M^{\star}$ galaxies have taken place
during a wide redshift interval $1<z<4$ for $q_0$=0.5, which becomes $1<z<3$ for
$q_0$=0.15. There seems to be 
a tendency for lower-mass ($M<5\ 10^{10}\ M_{\odot}$) systems
to have their bulk of SF protracted to lower redshifts. Our estimated galactic
masses, for a Salpeter IMF, are found in the range from a few $\sim 10^{9}\ M_\odot$
to a few $10^{11}\ M_\odot$ already at $z\simeq 1$.  So the bright end of the E/S0
population is mostly in place by that cosmic epoch, with space densities, masses and
luminosities consistent with those of the local field E/S0 population. 
We argue that the
strong decrease of the comoving mass density of early-type galaxies found by some
authors already by $z\simeq 1$ might be due to improper color classification, since
these objects are usually found to display blue young populations mixed with old red
stars. Instead, what distinguishes the present sample is a remarkable absence of
objects at $z>1.3$, which should be detectable during the luminous star-formation
phase expected to happen at these redshifts.  Obvious solutions are {\it a)} that
the merging events triggering the SF imply strongly perturbed morphologies which
prevent selecting them by our morphological classification filter, or {\it b)} that
a dust-polluted ISM obscures the (either continuous or episodic) events of
star-formation, after which gas consumption (or a galactic wind) cleans up the
galaxy. We conclude that the likely solution is a combination thereof, i.e. a set of
dust-enshrouded merging-driven starbursts occurring during the first few Gyrs of the
galaxy's lifetime. While our main conclusions are moderately dependent on the
assumed value of $q_0$, an open universe is favored in our analysis by the match of
the K-band local luminosity functions with the observed numbers of faint distant
galaxies. Two sources of uncertainty in our analysis -- i.e. the possible presence
of a background cluster or group at $z\sim 1$ in the HDF (possibly contaminating the
z-distribution), and the lack of a complete spectroscopic identification -- are
shown not to likely affect our main results. In any case, they will be reduced soon
by new observations in the Southern HDF and by deep spectroscopic surveys with large
telescopes.

\end{abstract}

\keywords{galaxies: ellipticals -- galaxies: ISM -- dust, extinction
infrared: galaxies }

\section{INTRODUCTION}

Early-type galaxies have been studied in great detail in rich clusters
up to redshift one and above (see Stanford, Eisenhard \&
Dickinson, 1998), with the basic
result that passive evolution in luminosity and no evolution in mass is
ruling them to at least $z\simeq 1$. Old ages and an early coeval epoch of
formation are then implied by these observations.  It is still unclear, 
however, how much the rich-cluster environment is representative
of the general population. Indeed, contradictory results have been
reported about the evolutionary properties of early-type field galaxies.

The analysis by Im et al. (1996) of a sample of elliptical galaxies in the HST
Medium Deep Survey seems to indicate that these galaxies share similar
properties with the cluster objects, a result based however on a sample of
376 galaxies with only 24 spectroscopic redshifts and photometric 
redshifts based on (V--I) colors. Similarly, Lilly et al. (1996) find that the luminosity function of red galaxies in the Canada-France
Redshift Survey (CFRS) does not exhibit significant changes in the redshift
interval $0.2\leq z \leq1$. These observations imply an early
epoch of formation for the elliptical galaxies.

These conclusions have been questioned by Kauffmann, Charlot, \& White (1996)
(see also Baugh, Cole, \& Frenk, 1996), who claim evidence for a decrease 
in the mass 
density already at $z\leq 1$ for early-type galaxies in the CFRS, 
these too selected from V--I colors.

An exceedingly deep and clean view of the field galaxy populations to high
redshifts is provided by a long integration of HST in the so-called {\it
Hubble Deep Field} (HDF, Williams et al., 1996).  Though having provided
relevant constraints on galaxy formation and evolution and about the
epoch of production of metals (e.g. Madau et al.\
1996; Connolly et al.\ 1997; Sawicki, Lin, \& Yee, 1997), 
the interpretation of these data alone is made difficult by the
relatively short selection wavelengths (essentially those
of the U, B, V, I bands), which
imply strong evolutionary and K-corrections as a function of redshift
(Giavalisco, Steidel, \& Macchetto, 1996). 
These may be particularly severe for
the early-type galaxies, because of their quickly evolving optical
spectra and extreme K-corrections (e.g. Maoz, 1997).

Another unsolved problem specifically affecting short wavelength observations
concerns the possible effects of dust present
in the line-of-sight to the object: even small amounts may seriously affect
spectra corresponding to rest-frame far-UV wavelengths.
Various analyses attempting to solve this issue have compared the observed
UV spectra with templates of young dusty galaxies (e.g. Meurer et al.\ 1997).
It is evident, however,
that even small variations in the dust properties and in the assumed spectral
templates imply widely discrepant predictions for the extinction corrections
(published estimates range from less than 1 to more than 3 magnitudes of 
extinction in high-z galaxies). Any inferences
about the history of star-formation are correspondingly uncertain. 

We try in this paper an alternative approach to the past history of galaxies
based on a thorough photometric and morphological analysis of a complete
K-band selected sample of elliptical and S0 galaxies in the HDF.
The selection in the K infrared band helps to overcome most of
the above problems of optical selection, in particular minimizes the effects
of K- and evolutionary corrections and of extinction by any residual
intervening dust. Another crucial advantage of using near-infrared data
is that the integrated fluxes at these wavelengths are contributed by
all (low-mass) stars dominating the baryonic content of a galaxy.

Various deep integrations have been performed in the HDF at near-IR
(Cowie et al.\ 1996; Dickinson et al. 1997) and mid-IR wavelengths
(Rowan-Robinson et al. 1997; see also 
Aussel et al. 1998), which, combined with
the extreme quality of the morphological information and the very good
spectroscopic coverage, make this area unique, in particular for the
investigation of early-type galaxies outside rich clusters.

We concentrate here on a (morphologically selected)
sample of 35 early-type galaxies. Our interest for this sub-population rests
on its homogeneous morphological properties, indicative of a
purely stellar emission (hence requiring a relatively simple
modelling), and on the expected old ages (by comparison with their rich cluster
analogues). Altogether, this class of objects is expected to inform about
the high-redshift side of the star-formation history, which has perhaps been 
only partly sampled by direct optical selection via the {\it drop-out} 
technique. In our approach, 
the inferred evolutionary history of stellar populations is essentially
unaffected by the mentioned problems related to dust extinction.

Section 2 of this paper provides details about the selection of the sample,
the adopted procedures in the photometric and morphological analyses, and
statistical tests of completeness carefully dealing with 
observational limits on the total flux and surface brightness. 
In Section 3 we analyze physical properties of galaxies as
inferred from the morphology and from fits to the broad-band UV-optical-IR
spectra. Section 4 addresses the statistical properties of the sample
(counts, redshift distributions, identification statistics), and compare them
with those of local samples of early-type galaxies. 
We discuss, in particular, evidence that the redshift distribution
breaks above $z\gtrsim 1.3$.
Various interpretations of these results are discussed in Section 5, 
where the effects of repeated merging events, probably in the presence 
of a dust-polluted medium, are considered. 
This information is used in Section 6 to constrain the history of star 
formation of stellar populations, and to compare it with other published 
estimates.
Our main results are finally summarized in Section 7.

We adopt $H_0=50\ Km/s/Mpc$ throughout the paper. The analysis is made for
two values of the cosmological deceleration parameter, $q_0=0.5$ and 0.15, assuming
zero cosmological constant $\Lambda$. Note that the effects of increased time-scales 
and volumes of an open universe might be also obtained in a closure world model
with a non-zero $\Lambda$.

\section{SAMPLE SELECTION AND PHOTOMETRY}

Dickinson et al. (1997) obtained deep near-infrared images of the $HDF$ 
with the IRIM camera mounted at the KPNO-4m Telescope. The
camera employs a 256x256 NICMOS-3 array with $0''.16$/pixel, but the
released images are geometrically transformed and rebinned (with
appropriate pixel weighting) into a 1024x1024 format. 
IRIM has observed the same area in the J, H and K filters, for a
total of 12, 11.5 and 23 hours, respectively. Formal
5-$\sigma$ limiting magnitudes for the HDF/IRIM images, computed
from the measured sky noise within a $2''$ diameter circular aperture,
are $23^m.45$ at J, $22^m.29$ at H, and $21^m.92$ at K, whereas the
image quality is $\sim 1''.0$ $FWHM$.

Our galaxy sample has been extracted from the HDF/IRIM K--band image 
through a preliminar selection based on
the automatic photometry provided by SExtractor (Bertin and Arnouts 1996).
It is flux--limited in the K--band and it includes only galaxies whose
morphology strongly suggests an {\it early--type} classification. The measure of
magnitude is based on fixed  2.5$''$ aperture photometry (which best accounts for
the seeing of the K image) and applying a stellar correction for the outer part.

\subsection{Monte-Carlo tests of completeness}

In order to set the appropriate value of the magnitude limit $K_L$ for
inclusion in our sample, we have produced a synthetic frame containing
a 10$\times$10 grid of toy galaxies with $r^{1/4}$ luminosity profiles,
having total $K$ magnitudes and effective radii $r_e$ spanning the
ranges $19^m.00\div 21^m.25$ (step $0^m.25$) and
$0^{\prime\prime}.08\div 0^{\prime\prime}.80$ (step
$0^{\prime\prime}.08$), respectively. These limits were chosen to
provide a full characterization of the performances of SExtractor
(i.e. identification capability and magnitude estimate of galaxies) in
the critical range of $K$ magnitudes and for values of the true
effective radius typical of the $HDF$ ellipticals (Fasano and
Filippi~1998, hereafter FF98; see also Fasano et al.~1998, hereafter
FA98).

We are particularly interested in evaluating the probability of
detection, as well as possible biases in the magnitude estimation, as
a function of the magnitude itself and of $r_e$ (i.e. of the `true'
average surface brightness $<\mu_e^K>$, and, after all, of the
redshift).  We have convolved this synthetic frame using a $PSF$
derived by multi--gaussian fitting of the few stars included in the
$K$--band image. Then, a bootstrapping procedure has been carried out
in order to produce 10 different images mimicking the 'true' image
noise. These images have been processed with SExtractor to get,
for each toy galaxy, ten different SExtractor estimates of the total 
$K$ magnitude, and then the average magnitude $<K_{SEx}>$, the standard
deviation $\sigma_{K_{SEx}}$ (or the difference $\Delta K =
<K_{SEx}>-K_{true}$).

From these simulations of galaxies having $r^{1/4}$ luminosity profiles
we conclude the following. {\it a)} For the
whole range of tested magnitudes and radii, the fraction of detections
is equal to unity up to $<K_{SEx}>\simeq 21^m.4$ and $<\mu_e^K>\simeq
22^m.0$. {\it b)} 
The magnitudes estimated by SExtractor are systematically fainter
than the true magnitudes, the bias mainly depending on the
galaxy surface brightness.

Figure 1a shows that, if we consider only galaxies with
$<K_{SEx}>\lesssim 20^m.7$ ($K_{lim}$ hereafter), the standard
deviation of the SExtractor magnitude estimates is less than $0^m.05$
($\sigma_{max}$ hereafter).  Figure 1b reports $\Delta K$ as a
function of $<\mu_e^K(SEx)>$ for the sub--sample of toy galaxies with
$<K_{SEx}>\lesssim K_{lim}$. It shows that $\Delta K$ systematically
increases at increasing $<\mu_e^K(SEx)>$, approaching the maximum value
$\Delta K_{max}\sim 0^m.5$ for $<\mu_e^K(SEx)>\simeq 22^m.0$. The solid
line in Figure 1b is a polynomial fit to the data:

\begin{equation}
\log(\Delta K) = -0.830 + 0.147\times<\mu_*^K> + 0.019\times<\mu_*^K>^2
\end{equation}

\noindent where $<\mu_*^K> = <\mu_e^K(SEx)> - 19$.

The plots in Figure 1 provide a straightforward indication of the
strategy to be used in order to obtain a complete flux limited sample
of early--type galaxies from the $IRIM$ $K$--band image using
SExtractor. The procedure consists of the following steps: 1) to
produce a catalog of objects with $K_{SEx}\lesssim K_{lim}=20^m.7$;
2) to check the morphology of each object on the WFPC2 frames,
removing stars and late--type (or irregular) galaxies from the sample;
3) to exploit the high resolution and depth of the $V_{606}$ and/or
$I_{814}$ images to derive the effective radii of the galaxies and use
these values of $r_e$ to compute $<\mu_e^K>$, assuming a roughly
constant color profile; 4) to apply to the $K_{SEx}$ magnitudes the
statistical corrections given in the previous equation; 5) to include
in the final sample only galaxies with corrected magnitude less than
or equal to $K_L = K_{lim} - \Delta K_{max} - \sigma_{max} = 20^m.15$.

\subsection{The morphological filter}

The most delicate steps in the previous scheme are those concerning
the morphological analysis and the evaluation of the effective radii
(points 2 and 3 of the previous list). They are widely discussed in
FF98 and FA98 and we refer to these papers for details. Here we wish
to summarize the 'flow--chart' of the morphological filtering for the
present sample. The SExtractor automatic photometry of the $IRIM$
$K$--band image produced a preliminar sample of 109 objects with
$K_{SEx}\leq 20^m.7$ (first point of the previous
selection scheme). All these objects were examined with the {\it
IMEXAM--IRAF} tool to produce a first (conservative) screening against
stars or late-type and irregular galaxies, resulting in a temporary
list of 47 {\it early--type} candidates. 
Our quantitative morphological classification filter assumes a dominant
$r^{-1/4}$ profile for the bulk of the light distribution in the galaxy.

For 29 galaxies in this list
the surface photometry in the ST--$V_{606}$ band is available in FF98
and the total AB magnitudes in the four $WFPC2-HDF$ bands are given in
FA98, together with the equivalent effective radii derived after
deconvolution of the luminosity profiles. Four more galaxies included
in both our and FF98 samples, were not analyzed by FA98. For these
galaxies we computed optical AB magnitudes and effective radii according 
to the prescriptions given in FA98. The remaining 14 galaxies in the
temporary list are not in the FF98 sample, since in the $V_{606}$ 
band they did not satisfy the selection criteria adopted in order to 
secure a reliable morphological analysis. We performed the detailed
surface photometry of these galaxies in the $I_{814}$ band (where they 
show a better S/N ratio), producing luminosity and geometric profiles 
of each galaxy. From this analysis, 3 objects were recognized to
be '{\it disk-dominated}' galaxies (likely $Sa$), whereas two more objects 
showed peculiar or unclassifiable profiles. These galaxies were excluded
from the temporary list. 

The remaining 9 galaxies with surface photometry 
in the $I_{814}$ band were analyzed following the same procedure described 
in FA98 to derive the total AB magnitudes in the optical bands, 
as well as the '{\it true}' (deconvolved) equivalent effective radii.
Figure 2 shows the luminosity profiles of these galaxies derived from 
the surface photometry in the $I_{814}$ band.

\subsection{The final sample}

Altogether, of the 47 candidate E/S0 galaxies with $K_{SEx}\leq 20.7$, 33 were
already classified as E/S0 by FA98 and FF98, while 9 are confirmed as E/S0
by the present paper, for a total of 42 objects in the incomplete sample at
$K_{SEx}\leq 20.7$. To ensure a highly reliable completeness limit, following
steps 4 and 5 of the selection scheme discussed  in Sect. 2.1, 
we used equation (1) to correct the $K_{SEx}$ magnitudes of the 42
galaxies in the temporary sample. In this way we 
obtain our final sample of 35 galaxies with corrected magnitude $K\leq 20^m.15$ over the
HDF (3 WF + 1 Planetary Camera) area of 5.7 square arcmin. Some basic data 
on the sample are listed in Table 1. 

The $H$ and $J$ magnitudes listed in the Table have been obtained
running SExtractor on the corresponding $IRIM$ images and
accounting for the expected biases with the same procedure adopted for
the $K$ magnitudes (simulations of toy galaxies, convolution with proper
$PSF$, noise bootstrapping and correlation between bias and average
surface brightness). The corresponding equations are:

\begin{equation}
\log(\Delta H) = -0.903 + 0.288\times<\mu_*^H> - 0.0158\times<\mu_*^H>^2
\end{equation}

\begin{equation}
\log(\Delta J) = -1.014 + 0.162\times<\mu_*^J> + 0.0140\times<\mu_*^J>^2
\end{equation}

\noindent 
where $<\mu_*^H> = <\mu_e^H(SEx)> - 20$ and $<\mu_*^J> = <\mu_e^J(SEx)> 
- 19$.

For 15 galaxies in Table 1 spectroscopic redshifts are available in the
literature or in the WEB (see notes to Table 1), while for the remaining 
20 objects a
reliable estimate of $z$ is obtained from fits of the optical-IR
broad-band spectra (see Sect. 3.2).

The last column in Table 1 gives some information on the surface photometry.
In particular, the symbol (814 Ph.) indicates galaxies whose surface 
photometry has been obtained in the $I_{814}$ band, while the symbols 
(T1), (T2) and (T3) indicate galaxies which, according to FF98, belong 
to the `{\it Normal}', '{\it Flat}' and '{\it Merger}' class, respectively. 
These attributes refers to the classification by FF98 
(see also Fasano et al. 1996),
based on the luminosity profiles, of {\it early--type} galaxies 
in the $HDF$: {\it 1)} the '{\it Normal}' class, for which
the de~Vaucouleurs law is closely followed down to the innermost isophote 
not significantly affected by the $PSF$; {\it 2)} the '{\it Flat}' class, 
characterized by an inward flattening of the luminosity 
profiles (with respect to the de~Vaucouleurs' law) which cannot be 
ascribed to the effect of the $PSF$ ; {\it 3)} the '{\it Merger}' class, 
in which isophotal contours show the existence of complex 
inner structures (two or more nuclei) embedded inside a common envelope, 
which roughly obeys the $r^{1/4}$ law.

The completeness of the sample has been evaluated from numerical simulations 
as described in Sect. 2.1. 
We have not used the classical $V/V_{max}$ test to this purpose,
since it is equally sensitive to departures from spatial homogeneity
than to completeness. The question of the distribution of objects in the
space-time will be addressed in Sect. 4 below.


\section{ANALYSIS OF THE BROAD-BAND SPECTRA AND OF THE SURFACE BRIGHTNESS
DISTRIBUTIONS}

The surface photometry of the images of our sample galaxies has shown that 
for the large majority of them 
there are no morphological signatures of the presence of dust,
e.g. obscured lanes across the galaxy, asymmetric brightness distributions
or profiles deviating from the $r^{-1/4}$ over the bulk of the galaxy
(see however Witt, Thronson, \& Capuano, 1992, for a cautionary remark about the
latter point).
Only three galaxies belong to the morphological class T3, where on-going
star-formation, possibly enshrouded by 
dust, is likely to have a role. Four more objects 
belong to the class T2, for which any dust, whenever present, should most probably 
be confined to the inner core of the galaxy, hence should not affect the 
global emission. Five out of 7 of the morphologically peculiar objects have blue (class [a], as discussed in Sect. 3.2) spectra.

An obvious way to test for the presence of dust would be to look for its
re-radiation at infrared wavelengths. The Infrared Space Observatory has
observed the HDF in two broad-band filters centered at 6.7 and 15 $\mu m$.
While the former, for redshifted objects, is dominated by photospheric 
emission of old stars, the latter is contributed by very hot dust and
transiently heated molecules or very small dust particles present in the ISM.
Mann et al. (1997) and Goldschmidt et al. (1997) report catalogues of sources
down to rather conservative flux limits. At 15 $\mu m$ only one ISO source is
in common with our list (ID 2-251-0), a galaxy with a point-source evidencing
in the U and B, probably an AGN. Though a conclusion has to wait for a 
refined analysis of the ISO observations, aimed at deeper flux limits 
(Aussel et al. 1998, Desert et al. 1988), 
the available information seems to confirm       
that there is little room for dust emission by our E/S0 galaxies. 

The present analysis will take advantage of this simplified behavior for the 
bulk of our objects,
by allowing modelling of galaxy spectra as the integrated emission of purely 
stellar populations. The inclusion of the additional effect of dust extinction,
whenever present, would have substantially weakened our conclusions, since
degenerate sets of solutions could have been possible, implying
uncertain estimates of the photometric redshift and of the galaxy age.

\subsection{The interpretative tool: a model for the evolutionary
spectral energy distributions of the integrated starlight}  

The UV-optical-near/IR Spectral Energy Distributions (SED) for all
galaxies in the sample have been fitted with synthetic spectra based on
a model briefly summarized here. A general framework for this modelization 
can be
found in Bressan, Chiosi, \& Fagotto (1994), while a critical assessment of problems
and limitations is discussed by Charlot, Worthey, \& Bressan (1996).

The code follows two paths, one describing the chemical evolution 
of the galaxy's ISM, the other
associating to any galactic time a generation of stars with the
appropriate metallicity, and adding up the contributions to the integrated
light of all previous stellar populations.

The chemical path adopts a Salpeter IMF with a lower limit $M_l= 0.15\ 
M_\odot$, and a Schmidt-type law for the
Star Formation Rate (SFR): $\Psi(t)=\nu \, M_{g}(t)^{k}$, where $\nu$ is the
SF efficiency. A further parameter is the time-scale $t_{infall}$ for the 
infall of the primordial gas. The library of isochrones is 
based on the Padova models, spanning a
range in metallicities from $Z=0.004$ to 0.05 (i.e. from 0.2 to 2.5 solar).
This range is appropriate for stellar populations in local early-type 
galaxies, which have solar metallicity on average 
(e.g. Grillmair et al. 1996).

The isochrones are modified to account for the contribution by dust-embedded 
Mira and OH stars during the AGB phase (Bressan, Granato \& Silva 1998), when
circum-stellar dust reddens the optical emission and produces an IR bump
at 5-20 $\mu m$. This IR emission from AGB stars is important for
galactic ages of 0.1 to a few Gyrs.

A number of evolutionary patterns for the time-dependent SFR $\Psi(t)$
have been tried to reproduce the galaxy SEDs, to estimate the photometric
redshifts for galaxies lacking the spectroscopic identification, and to
fit the global statistical properties of the sample.
However, we will mostly refer in the following to two paradigmatic evolution 
cases.

The first model (henceforth Model 1) reproduces a classical scheme for the formation
of ellipticals (e.g. Larson 1974), i.e. a huge starburst on short
timescales, expected to occur at high redshifts.
In our approach the star-formation has a maximum at a galactic age of
0.3 Gyr and continues at substantial rates up to 0.8 Gyr, after which it
is assumed that the input of energy by stellar winds and supernovae
produced a sudden outflow of the ISM through a galactic wind, stopping
the bulk of the SF. The evolution at later epochs is mostly due to
passive aging of already formed stellar populations. 
This evolution pattern is achieved with the following choice of the
parameters: $t_{infall}=0.1$ Gyr, $k=1$, $\nu=2\ Gyr^{-1}$. 
The precise time dependence of the SF rate is reproduced in Figure 3 (dotted line).
The redshift for the onset of star-formation $z_F$ in the galaxy
is a free parameter.

Our second model (henceforth Model 2) 
gives up the concept that the stellar populations in field
early-type galaxies are almost coeval, and assumes instead that the
star-formation lasts for a significant fraction of the Hubble time.
This is obtained with the following parameters:
$t_{infall}=1$ Gyr, $k=1$, $\nu=1.3\ Gyr^{-1}$. 
The corresponding SF law has a broad peak at 1.4 Gyrs but goes on at
substantial rates for a couple Gyrs more. For a value of the
baryonic mass of $M=10^{11}M_\odot$ (as typical for our sample galaxies,
see below), this brings to an energetic unbalance in the ISM followed by
a galactic wind at 3.1 Gyrs (see continuous line in Fig. 3 
for the detailed evolution of the SF). 

This adopted SF law is clearly an over-simplified
picture of a process which has been likely more complex. In particular,
a protracted SF, as implied by the second model, is likely to have been occurred 
through a set of successive starbursts (e.g. due to mergers or strong dynamical interactions). Since we are dealing with the integrated emission of all
stellar generations, there is virtually no difference, as for the broad-band
spectral appearance in the after-burst phase, between a set
of starbursts occurring over 3 Gyrs and a continuous SF during this period.

To provide these simplified schemes with more flexibility, we have allowed
a residual star-formation $\Psi_0$ at a basal level to be added to the flux
emitted by the passively evolving populations. 
This allows to fit the spectra of the four bluest objects, and to improve
the fits at the shortest (U, B) wavelengths for additional galaxies.
Since typical values for  $\Psi_0$ are much less than 1 $M_{\odot}/yr$
and this residual SF does not contribute significantly to the mass and
energetics,
this parameter is used only in the spectral fits of Fig. 4 below and
never more in the subsequent analyses.

$\Psi_0$ is then a second free 
parameter. The baryonic mass $M$ and the photometric redshift (for cases
where it is needed) are further parameters in the spectral fitting procedure.

\subsection{Fitting the galaxy broad-band spectra: evaluation of the
photometric redshifts and of the galactic ages} 

\subsubsection{The photometric redshifts } 

Figure 4 is a collection of the observed spectra for the 35 sample galaxies. 
The figure displays essentially two kinds of spectral behaviors: 
class {\it (a)} spectra, which are rather blue at all 
wavelengths, dominated by young stellar populations (10 objects
out of 35); class {\it (b)} spectra, with overall redder properties, 
displaying a typical two power-law behavior, with a break at $\lambda
\simeq 0.4\ \mu m$ in the rest-frame spectrum in correspondence of the 
Balmer decrement (25 sources). 
No objects, even those at the lowest redshifts, are found to
display very red colors, as would be expected for a very old stellar
population dominating the spectrum. This result is consistent with 
a statistical study by Zepf (1997) revealing a lack of very red objects in 
the field.

A crucial step in our analysis is the evaluation of redshifts
from broad-band spectral fitting
for sources lacking a spectroscopic identification. The uncertainties 
related with this estimate have been discussed by many authors (see
in particular Connolly et al. 1997; Hogg et al. 1998),
with the general outcome that the inclusion of near-infrared data
in the analysis (added to the 4 HST bands) makes the redshift
estimate quite more reliable.

With respect to these analyses we benefit here of various advantages.
The first one is due to the lack of evidence for dust in our objects,
which breaks down at least one possible degeneracy in the parameter space
(that is estimating a lower photometric z from a dust-reddened template).
The second one is that most of the observed spectra have rather 
homogeneous and monotonic behaviors as typical of early-type galaxies,
with an easily discernible Balmer feature.
A third potential advantage, as verified "a posteriori", 
is given by our relatively bright K-band selection, which
tends to select objects only up to moderate redshifts ($z<1.5$, see Sect.
4-5) and to exclude very high-z galaxies, whose redshift estimate would be
quite more uncertain because of the strong evolutionary corrections in the 
template spectra.

More specifically, 
for class {\it (b)} sources lacking spectroscopic redshifts (i.e. 15 galaxies
in total)
the estimate of the redshift from spectral fitting is quite robust, thanks
to a well-defined spectral break corresponding to the Balmer decrement,
typically occurring at $\lambda \sim$ 0.6 to 1 $\mu m$. 
The combined use of the
3 near-IR and the 4 HST bands, with very small photometric errors, allows
an accurate determination of the 4000 A break from a model-assisted spectral 
interpolation.
For class {\it (a)} spectra without spectroscopic redshifts (5 more galaxies),
there is still clear evidence for a 4000 A break, which is however less
precisely characterized in some instances. In these cases the photometric 
redshift may turn out to be significantly more uncertain.

Figure 5 summarizes a comparison of our redshift estimate based on broad-band
spectral fitting with the actual measurement from mid-resolution spectroscopy, for
the 15 galaxies in our sample with this information. The errorbars associated with
the photometric estimate are conservative $\sim$95\% confidence intervals based on
$\chi^2$ fitting using Model 2 as spectral template. In only one case 
(object 2-251-0) the photometric
measure is significantly discrepant with respect to the spectroscopic one, but
this happens for the galaxy containing a nuclear point-like source, 
which affects the match of HST and near-IR photometry.
Fig. 5 shows that our process exploiting 7-band spectral data is 
overall quite reliable and that no systematic effects are present.
On the other hand, it is clear that the 15 galaxies with optical spectroscopy 
are not randomply sampled from our source list, as shown in particular by Fig. 
7 below: they tend to be lower-redshift, bluer objects, with SF protracted
to recent cosmic epochs. Such spectral behaviour has encouraged a spectroscopic
follow-up, while the redder higher-z ones will require a substantial dedicated
future effort.

Altogether, we estimate that the errors associated with redshifts
evaluated from class {\it (b)} spectra should not typically exceed $10\%$
in $z$,
while a conservative estimate may be closer to $20\%$ for class {\it (a)}
spectra (see also Connolly et al. and Hogg et al. for similar conclusions).
None of our results will be significantly affected by these uncertainties.

\subsubsection{Evaluation of the galactic ages } 

For all sample galaxies, but for object 4-727-0, we found acceptable fits 
with Model 2. These solutions were found 
by optimizing the 3 (4) free parameters discussed in the previous Section.
The spectral solutions illustrated in Fig. 4 (whose parameters are reported
in Table 2) refer to the case $q_0=0.15$. Consistent solutions were also 
found for Model 2 in a closure world model 
(see again Table 2 for the corresponding best-fitting parameters).

The failure of Model 1 to reproduce the data is illustrated in Figure 6,
which compares the broad-band spectrum of a typical galaxy in the sample
with two 
predictions of the model at varying $z_F$. The continuous curve is
the spectrum predicted by Model 1 with $z_F=5$ and $q_0=0.15$. With this choice 
of the parameters, this corresponds to a stellar population beginning 
formation at z=5 and ending it at z=3.6 (hence observed 5.5 Gyrs after the 
end of the SF in the source frame): the prediction is clearly far too 
red both in the optical and even in the near-IR. The dashed curve
corresponds to the same model with $z_F=1.3$, and the additional contribution
of ongoing SF by 0.2 $M_{\odot}/yr$ to better reproduce the U band flux.
If the overall fit is better, there is still a significant excess flux
in the observed B band with respect to the model spectrum. We have found this
excess, as well as sometimes one in the V band, to be a general
characteristic of the observed spectra: 
even for the reddest galaxies in the sample it seemed difficult to reproduce the spectra  with stellar populations
formed during a relatively brief time interval (as in Model 1), even including the additional contribution of ongoing star-formation.

The better performance of Model 2 is interpreted by us as indicating a rather
protracted star formation activity within each galaxy (typically 3 Gyrs in the 
model), rather than a single short-lived starburst (even one occurring at low
redshift). A qualitatively similar result is achieved 
by FA98 analyzing a sample of HDF early-type galaxies 
selected in the $V_{606}$ band.

The second conclusion concerns the epoch when the stellar 
formation has taken place in the typical sample galaxy. Figure 7 is a plot of
the redshift corresponding to the peak of the star-formation versus the
mass in baryons, according to our best-fits obtained with Model 2. 
Panels (a) and (b) refer to the two solutions with $q_0=0.15$ and $q_0=0.5$.
There are clearly two zones of avoidance in the figure: low-mass objects
($M<2\ 10^{10}\ M_{\odot}$) have peak SF confined to $z\lesssim 1$, while 
the high-mass ones 
appear to form stars mostly at $z\gtrsim 1$. This is partly an artifact
of the magnitude limits (in the former case) and of the small sampled
volume combined with the low space density of very massive galaxies 
(in the latter). While only a careful examination of all selection effects
will allow to obtain unbiased estimates of the cosmic SF history
from this dataset (see Sect. 6 below), {\it it is clear from Fig. 7 that the 
redshift interval of
z=1 to 2 (1 to 3 for $q_0=0.5$) corresponds to a very active phase of SF
for the sample galaxies}.

To translate this into a constraint on the age for the typical
early-type galaxy in the field, we report in Figure 8 the rest-frame V--K
and B--J colors as a function of redshift, compared with the predictions of 
single stellar populations with solar metal abundances. The rest-frame B--J
colors are computed by interpolating the galaxy observed spectra using the 
best-fit models of Table 2, while the V--K's require substantial extrapolation
at the longer wavelengths, hence are to be taken with some care.
The figure shows that the typical ages range from 1.5 to 3 Gyrs, rather
independently from the redshift (if any, there is quite a marginal tendency 
for the higher-z galaxies to display bluer colors).
For comparison, the average colors of local galaxies are V--K$\simeq$ 
3.2--3.3, quite significantly redder. 
This difference is probably enhanced by a bias induced by the flux limit 
in the K-selected sample, emphasizing relatively bluer objects observed
close to their main event of star-formation. Again this will be subject of
further inspection in the next chapters.

Another check of this is reported in Fig. 6, where the broad-band
spectrum of the typical galaxy is compared with that of the local
early-type M32 (F. Bertola, private communication). The present
interpretation of M32's blue light is that it may be dominated by the
HR turn-off population
of a $\sim 4$ Gyrs old stellar population of about solar
metallicity (O'Connell 1986), whose bright RGB and AGB stars have been
recently resolved (Freedman 1992, Grillmair et al. 1996). Older ages
might fit only for a metal-deficient dominant population (Renzini \& Buzzoni
1986). Our typical galaxy is bluer than  M32 and, assuming  a solar 
metallicity, it turns out to be younger than 4 Gyrs.

The previous considerations have impinged upon a well-known problem when 
dating stellar populations from broad-band spectral analyses, i.e. the 
degeneracy between age and metallicity. An estimate of the age needs 
a guess on the metallicity.
As already discussed, the very regular morphologies and the lack of evidence
for the presence of dust in the large majority of our sample galaxies
suggest that they have already mostly completed their bulk of the SF 
and have metallicities comparable to those of the 
local counterparts (i.e. solar on average, see Carollo \& Danziger 1994). 
If true, then a conclusion seems unavoidable, that their typical age is close to 
1.5 to 2 Gyrs on average, at the observed redshift. 

Figure 9 illustrates the effects of drastic changes in the metallic 
content of the stellar populations on the rest-frame colors. 
The two horizontal lines in both panels bracket
the typical colors for the sample galaxies. 
Large age differences are found as a function of the metallicity for 
a given color and for old stellar populations. 
On the other hand, {\it since our observed colors are rather blue, 
the uncertainty due to the unknown metallicity is moderate in absolute terms}.
For example, with reference to Fig. 9b, the typical observed color
B--J$\simeq 2.5$ would correspond to an age of 1 Gyr for 2 times solar,
to 2 Gyr for a 0.4 solar and 3 Gyr for a 0.2 solar metallicity.
The conclusion is that, {\it unless we accept that the observed galaxies 
are very significantly metal-deficient (which would be difficult to 
reconcile with observations of local objects), our age determinations should
not be seriously in error}.

\subsection{Constraints from the size vs. surface brightness distributions
}  

Further significant constraints onto the nature and evolutionary status
of our sample galaxies may be gained from a detailed analysis of their
morphological properties.

The study of the galaxy spectral shapes in the previous Sect. has 
indicated that intermediate (rather than old) ages, with large spreads, 
are typical. If so, then one would expect quite appreciable luminosity 
evolution due to the -- mostly passive -- aging of the stellar populations
from the redshift of the observation to the present time.
A way to check it, as discussed in FA98, would be to compare
the sizes and the average surface brightness (within the effective radius)
of distant galaxies with those of the local ones. 
Table 1 (columns 6 and 15) reports these data for all galaxies in our sample.

Figure 10 compares the same data with the mean locus ((thick continuos 
line) of the
relation between the effective radius $r_e$ and the corresponding average 
surface brightness in the V band $<\mu_e>_V$ (the Kormendy relation). 
In this figure the effective radii are expressed in Kpc,
while the surface brightness has been corrected (K- and evolutionary
corrections) according to Model 1.

There is a clear offset in Fig. 10, by $\sim 1.5$ magnitudes, with respect to
the locus representing local galaxies (see for more details on the latter
FA98 and Jorgensen et al., 1995): 
the surface brightness of distant galaxies is
more luminous on average than implied by Model 1 -- one in which, we remind,
the bulk of SF was completed by $z=3.5$.

Figure 11 shows that much better consistency is achieved with Model 2,
both in the V (panel [a]) and in the K bands (panel [b]).
The local Kormendy relation in the band plotted in Fig. 11b is taken from 
Pahre et al. (1995). 
In this case the evolution in luminosity between the redshift of the
observation and the present time is stronger (stars are younger on average),
with local and distant galaxies consistently tracing the same population,
as expected.

\section{STATISTICAL PROPERTIES OF THE COMPLETE SAMPLE: THE REDSHIFT 
DISTRIBUTION}

Though small, our K-selected sample has been tested with great care
in Sect. 2 for completeness and reliability in source selection,
and it is then suitable for a detailed statistical analysis.
In consideration of the high-quality morphological and photometric data
available, this study may provide an accurate characterization of the 
distribution of elliptical/S0 galaxies in the space-time. 

We consider here two statistical observables, the counts as a function of
morphological type and of limiting K magnitude, and the redshift distribution 
D(z).

Figure 12 is a collection of galaxy counts in the K and HK bands, splitted 
into two morphological components: the early-type galaxies (open squares), 
selected according to the criteria described in Section  2 (see Table 1), 
and late-type systems -- including spirals, irregulars, and starbursts 
(open triangles). 
The K band counts in the left-hand panel
 are derived from our sample discussed in 
Sect. 2 down to $K=20.1$ . The counts in the H+K band in the right-hand panel 
are derived from a sample by Cowie et al. (1996), including complete 
morphological information for HK brighter than 21.5.


The number counts for early- and late-type galaxies display rather different 
slopes at the faint end, with the early-types converging very fast at
$K>19$ and the late-types showing steadily increasing counts.
Similar results about the morphological number counts are reported by 
Driver et al. (1988), who find in particular a deficit of E/S0 at
$I_{814}>22$ compared with passively evolving models.

The distribution of the -- either spectroscopic or photometrically estimated -- 
redshifts for the 35 sample galaxies is reproduced as a thick line in Figure
13. The observational distribution shows monotonic increasing values
from $z$=0 to $z\simeq 1.3$, with a marked peak at $z\sim 1.1$. 

A remarkable feature in the distribution is 
a sudden disappearance of objects at $z\gtrsim 1.3$.
This absence may look surprising at first sight, taking into account that
the dominant stellar populations 
for objects observed at $z\sim 1$ are 1 to 3 Gyr old, whose luminosity
has then to increase, if any, at $z>1$.

Let us first discuss the significance of this apparent redshift cutoff, in the 
light of all selection effects operating in the sample. The first obvious
concern is our ability to identify in the K-band image
extended emissions corresponding to high redshift elliptical/S0's when going 
from the typical observed redshift $z\sim 1$ to larger cosmic distances.

Figure 14 illustrates the effect of increasing redshift on the average surface 
brightness and the effective radius, for two evolutionary paths
(both adopting $q_0=0.15$). 
The top line corresponds to a typical galaxy in our sample having 
$<\mu_e^K>=18$, $r_e=2$ Kpc (cfr. the distribution of values scaled to 
zero-redshift in Fig. 11b), and observed at $z$ ranging from 0 to 2.5.
The scaling with $z$ of the surface brightness has been calculated taking into
account all cosmological, K- and evolutionary effects. 
To be conservative, these were computed according
to Model 1 with a low formation redshift $z_F=3$. It turns out 
that the brightening of the stellar populations, while approaching the SF 
phase at $z>1.5$, counter-balances any cosmological and K-correction
dimming. In this case there is no significant dependence of the surface
brightness on redshift. A limiting situation is provided by the bottom curve
in Fig. 14, representing the lowest surface brightness galaxy in the sample
and luminosity evolution after Model 1 with a high formation redshift $z_F=6$. 
In this case the brightness dimming effects dominate and there is an appreciable
excursion in  $<\mu_e^K>$ as a function of $z$. 
In spite of this, in this case as well as
in all other more favorable ones, the faint object would be still be
detectable up to z=2.5
above the completeness limit, estimated by simulations in Sect. 2 to be  
$<\mu_e^K>=22$.
Note that the same conclusions hold for any more actively evolving models
(e.g. Model 2).

Altogether, the K-band image is sensitive enough to allow easy detection of 
even moderately or non evolving galaxies up to at least $z\sim 2.5$. 
Then the 
turnover in D(z) apparently occurring at much lower $z$ cannot be due
to limitations in detecting faint extended structures.

We investigate in the following how the other fundamental selection condition, that 
in the total flux ($K<20.15$), operates for different 
cosmological and evolutionary models.
To perform this exercise, a reliable local luminosity function (LLF)
of galaxies in the K band is required. Our adopted LLF is taken from
Gardner et al. (1997), which significantly updates previous
determinations. The separate contributions of early- (E/S0) and late-type 
(Sp/Ir) galaxies have been calculated using the optical LLFs 
splitted into various morphological types by Franceschini et al.
(1988) and translated to the K band with type-dependent B-K colours.

The K-band luminosity function of galaxies has been combined with 
various evolutionary models to predict sample statistics,
under the simplifying assumption of a galaxy mass function
constant with cosmic time.

We find that the shape of the LLF estimated by Gardner et al. 
can be more naturally
reconciled with the total faint K-band counts of Fig. 12 in an open universe with
$\Lambda=0$, 
while a closure one would require some "ad hoc" evolutionary prescriptions. 
This is because the latter has not enough volume at $z\sim 1$ to fill in the 
counts, given the space density of galaxies implied by the K LLF.
Then Model 1 for E/S0 galaxies (see Sect. 2.2), 
supplemented with a moderate evolution for Sp/Ir galaxies as in Mazzei
et al. (1992), could in principle provide a fair fit to the total
counts for $q_0=0.15$.

But a further crucial constraint is set by the observed $z$-distribution
D(z) of early-type galaxies in the K-HDF (Fig. 13).  From $z=0$ to $z=1.2$, 
D(z) is consistent with a (mostly passive) 
evolution in an open universe (with $q_0$=0.1-0.2) with zero cosmological 
constant. Again, a closure world 
model with $\Lambda=0$ would require a much larger comoving 
luminosity density of massive E/S0s at $z$=1 than locally.

Hence, there is no evidence, up to this epoch, of an evolution of the baryon
mass function. This is difficult to reconcile with a decrease of the mass 
and luminosity due
to progressive disappearance of big galaxies in favor of smaller mass 
units, as implied by some specifications of the hierarchical clustering 
scheme (e.g. Baugh, Cole, \& Frenk, 1996). 
The case for a strong negative evolution of the early-type population
(by a factor 2-3 less in number density at $z$=1; e.g. 
Kauffmann, Charlot \& White 1996)
does not seem supported by our analysis of the HDF.

Above z=1.2, however, both Models 1 and 2 (whichever formation redshift $z_F$ 
is assumed) have difficulties to
reproduce the redshift distribution reported in Fig. 13 and the identification
statistics in Figure 12. 
In particular it is not able to explain the large number of sources  
at $z=1$ followed by the rapid convergence above: the
prediction would be of a much more gentle distribution, with less a
pronounced peak and a large tail of galaxies (including typically half of
sample objects) observable above $z=1.2$. 
Lower values of $q_0$ would allow better fits of the observed D(z) up 
to $z=1.3$, but would also worsen the mis-fit at higher $z$.

A similar effect is observed in the
counts as a function of the morphological class (Fig. 12). Here again
sources identified as E/S0s show steep counts to K=19 and a sudden
convergence thereafter, while Model 1 would predict a much less rapid
change in slope. These problems remain forcing in various ways the 
model parameters.

\section{DUST EXTINCTION AND MERGING EFFECTS DURING A PROTRACTED SF PHASE}

We briefly discuss here possible explanations of the results obtained 
in Sections 3 and 4.
We look for solutions explaining the global
statistical properties of the sample (see Sect. 4) as well as the 
variety of the single galactic spectra (Sect. 3).

\subsection{Merging}

In the previous Sect. we have found that
at redshifts larger than 1.3 early-type galaxies suddenly disappear 
from the K-selected sample, while we would expect to observe them to much
higher $z$, given the predicted luminosity enhancement when approaching
the most intense phase of star formation. 
A comparison of Figures 7 and 13 suggests that there may be a relation 
between the redshift interval where such a disappearance occurs and the one
bracketing the major episodes of star formation, 
as evaluated from the analysis of the stellar populations in distant galaxies.
It is clear from Fig. 7 that the interval $1<z<3$ is where the peak 
of star formation occurs for the bulk of our sample galaxies, and this 
roughly coincides with the redshift interval where virtually no objects 
are found in our sample.

As discussed in Sect 3.,
there is not only a substantial spread in the formation epochs $z_F$
of the sample galaxies (see Table 2), but also
in single galaxy spectra there is evidence for the presence of stellar 
populations with widely different ages. A model with SF distributed
over a 3 Gyr time-interval was able to fit such spectra. 

So, {\it during this relatively long-lasting period of formation of stars,
our sample galaxies escape detection by our selection procedure. }
Two filters may operate in producing this:
either the morphological filter, or the K-band limiting flux, or possibly 
both simultaneously.

An obvious morphological transformation accompanying the formation of a sizable
fraction of stars is predicted by the merging picture for the formation
of early-type galaxies: two gas-rich systems undergo a deep dynamical
interaction, eventually bringing to a single merge product. During this process
there is a full reshuffling of both the gas component, with dynamically  
triggered SF, and also of the old stellar population, reflecting the 
rapidly changing gravitational field. Numerical simulations show
that an $r^{-1/4}$-law stellar distribution may originate in this way (
Barnes \& Hernquist 1996).
Given the spread of ages in our observed objects, it is likely that a few
to several of these merging events have taken place per single massive galaxy
during the first few Gyrs of the galaxy's lifetime.
Merging, mostly occurring at $z>1$, is then a very appealing interpretation
for our results.

Another effect of merging would be to lower the flux detectability during
the SF phase, due to the mass function rapidly evolving at $z>1$.
We did not attempt to quantify the possible effects of merging on the
visibility of the early evolution phases of E/S0 galaxies. We defer for an
exercise of this kind to dedicated treatments (e.g. Baugh, Cole, \& Frenk 
1996).

Another plausible reason of flux leakage during SF is related to
the extinction by dust in the medium where stars are forming. This dust
is very likely present, left over by preexisting generations of massive 
stars.  We spend the rest of this Section
in a more quantitative evaluation of the possible effects of dust.
An argument strongly suggestive of its occurrence will be given in Section 6.

\subsection{Dust effects during a prolonged SF phase in field ellipticals}

Dust is observed in significant, or even large, amounts in a wide variety
of objects at any redshifts. The most distant astrophysical sites
explored so far, the quasars at redshift 4 to 5, have shown to contain,
at that early epoch, dust amounts comparable with those of the most
massive galaxies today (Omont et al. 1997; Andreani, Franceschini, \& Granato, 1997).
But also distant radio galaxies, Lyman "drop-outs",
and damped $Ly\alpha$ galaxies (Pettini et al. 1997)
have invariably shown the presence of dust.
In several galaxies, among the infrared selected, IRAS has found 
that the dust re-processes a dominant fraction of the stellar UV radiation
into the infrared (the ultra- and hyper-luminous IR galaxies, see e.g. Sanders 
and Mirabel, 1996; Rowan-Robinson 1997).

Indeed, a single short-lived starburst may produce enough metals via
massive stellar outflows during the AGB and supernova phase. Recent ISO
observations (Lagage et al. 1996) have proven that dust grains are synthesized
almost simultaneously with the metals, as soon as they are made available 
to the medium.

A common interpretation of ultra-luminous IR galaxies is that, during a
merger or a close interaction, even moderate amounts of dust are spread
around to make an extended dusty core in the starburst. In these conditions
it is very likely that any major merger happens in a dust enshrouded medium,
several examples of which have been found at both low- (e.g. Arp 220,
NGC 6090, M82, see Silva et al. 1998) and high-redshifts (e.g. IRAS F10214;
HR10; see also Ivison et al. 1998).

We provide here a simplified description of the effects of dust during the
star-formation phase and test it on the statistical properties of our
sample as discussed in the previous Section.

The simplification we adopt is to assume, instead of a discrete set of 
successive merging-driven starbursts as would be a realistic physical
situation, a more continuous process of star-formation with a temporal
evolution following that of Model 2 in Figure 3 and occurring in a 
dust-enriched medium (we remind that in Model 2 the SF has a peak 
at 1.4 Gyr after beginning and keeps on until 3.1 Gyrs. We do not consider the 
effects of dust in Model 1, as it would only remove the $z>3$ sources and because
it is in any case inconsistent, see Sect. 3.2.2). 
Dust associated with the residual ISM 
extinguishes the light emitted by high redshift objects.
The idea is that the SF efficiency in field galaxies with deep dark-matter 
potentials is not so high to produce a galactic wind on
short timescales after the onset of SF. Hence an ISM is present in the 
galaxy for an appreciable fraction of the Hubble time, during which the ISM 
is progressively enriched of metals and dust. While the gas fraction
diminishes as stars are continuously formed, 
its metallicity increases and keeps the dust optical depth
roughly constant with time.

We have then complemented the Model 2 of Sect. 3.1 to account for
the effect of extinction and re-radiation by dust in star-forming
molecular regions and more diffused in the galaxy body during the active
phase of SF using a model by Silva et al. (1988). 
The model is precisely as described in Sect. 3.1 for the whole passively
evolving phase after the end of SF.

A detailed description of the code is given in Silva et al.
Here we summarize its basic features.
The amount of dust in the galaxy is assumed proportional to the residual gas
fraction and chemical abundance of C and Si, while the
relative amount of molecular to diffuse gas ($M_m/M_g$) is a model parameter.
The molecular gas is divided into spherical clouds of assigned mass
($5\ 10^{5}\; \rm{M}_{\odot}$) and radius ($16\ pc$).  
Each generation of stars born within the cloud progressively
escapes it on timescales of several tens of Myrs. The two parameters 
regulating this process (see eq. [8] in Silva et al. 1998) are the age 
$t_0$ of 
the oldest stars still embedded and the escape timescale $\alpha$.
The emerging spectrum is obtained by
solving the radiative transfer through the cloud.
Before escaping the galaxy, the light arising from young stars/molecular
complexes, as well as from older stellar generations, further interacts
with dust present in the diffuse gas component (the latter dominates the 
global galactic extinction at late epochs, when the SF rate is low).
The dimming of starlight and consequent dust emission are
computed by describing the galaxy as a spherically symmetric system
subdivided into volume elements, with radial dependencies of stars and gas 
described by a King profile with a core radius $r_c=200\ pc$,
as reasonable for early-type systems. 

In the general framework for the time evolution of the SF rate 
given by Model 2 (see Sect. 3.1 and the continuous line in Fig. 3), 
we found a self-consistent 
solution for the dust extinction and emission with the following 
parameter values: baryonic mass $M=10^{11}M_\odot$, (typical for our sample
galaxies), $M_m/M_g=0.3$, $t_0=0.1\ Gyr$ and $\alpha=100\ Gyr^{-1}$. After 
3.1 Gyr,
the balance of gas thermal energy vs. input from supernovae breaks, and
formally most of the small residual gas is then lost by the galaxy. 
After this event, a very low-level SF activity may keep on, due to either
partial efficiency of the wind, or to stellar re-cycling.
The average stellar metallicity of the remnant is roughly solar.

This scheme naturally accounts for the evidence  previously discussed 
in Sect. 3 of a substantial spread in the ages of distant field E/S0,
showing a combination of massive
amounts of old stellar populations and younger stars.

The same scheme, when convolved with a K-band LLF, 
successfully accounts for the statistical 
distributions in Figs. 12 and 13, in addition to other galaxy counts in the
K-band. Now the cutoffs of both D(z) at $z>1.3$ and of the counts for E/S0s 
at $K>19$ are reproduced as the effect of dust extinction during the SF phase.

We stress again that, though not physically unplausible, our treatment of
the SF as a continuous process during a substantial fraction of the
Hubble time at high-z is likely an over-semplification of a more complex
process characterized by a discrete set of SF episodes. However, there is    
virtually no difference between the two cases as far as the "observable"
later, passively evolving, phase is considered.

\section{THE STAR-FORMATION HISTORY OF EARLY-TYPE FIELD GALAXIES}

This Section is devoted to a quantitative evaluation  
of the  star-formation history of early-type galaxies in the field. 
Though it is not certainly the first time this is attempted,
our approach is substantially different and complementary, and brings some
significant advantages with respect to previous efforts
(see e.g. Lilly et al. Madau et al. 1996; Connolly et al. 1997; Madau, 
Pozzetti, \& Dickinson 1998).

Most of the previous analyses have concentrated on very large 
(e.g. thousands of objects for analyses dealing with the HDF itself or the 
CFRS) samples of galaxies without morphological differentiation and with 
sometimes very limited fractions (typically $<10\%$ if we exclude CFRS) of
spectroscopic identifications, directly
observed throughout the whole redshift range sampled by the very
sensitive optical images.

In addition to the admittedly uncertain redshift estimation via the Lyman
drop-out technique, a recognized difficulty inherent in these "direct"
evaluations of the cosmological star-formation rate is due to the
essentially unknown effect of dust. Given the poor knowledge of dust
properties, an extinction correction is subject to tremendous uncertainties
even for high-z galaxies with good optical spectra, which are in any
case a minority.

Our present approach attempts to by-pass all these kinds of problems by
confining the analysis to a relatively small sample (35 objects)
with extremely accurate information per object.
Such an information concerns high-quality photometric data over a large
wavelength interval ($0.3<\lambda <2.2\ \mu m$), a good fraction (43\%)
of spectroscopically confirmed redshifts, a very accurate morphological
information allowing to select a sample (that of early-types)
with likely homogeneous properties of star-formation.
The sample galaxies have been processed through a grid of synthetic
galaxy spectra, with the essential aim to date the stellar populations
present in the galaxy at the time of observation.

The selection waveband, the K band, has been chosen not only to minimize
the selection effects, due to the evolutionary and K-corrections, but
also to allow testing a wide range of stellar ages and masses, not
possible if only optical data were considered.
The basic uncertain factor when dating stellar populations, i.e. 
the metallicity of stars, was discussed in Sect. 3.2: our
conclusion was that, given the relatively blue colors of galaxies
at the time of observation, our results on the ages cannot be
drastically in error unless we consider highly non- (either super- or
sub-) solar metallicities. In particular our results should be correct
if our objects, having mostly completed their star formation, are 
characterized by the same solar metallicities which are typically observed 
in local galaxies.
The results of this best-fitting process are summarized in Table 2.

Our task here is to account for all reasonable selection effects operating
in the sample, to correct for incompleteness (that is for the part of the LF
unobserved because of the flux limit) and to build up a "population" 
history from those of the single galaxies.

This derivation of the SF history, though subject to the mentioned
modellistic uncertainties, should be minimally or not affected at all by the
dust extinction problem, as it deals with stellar populations during
a relatively late dust-free phase.

Our approach is otherwise the same as attempted by Lilly et al. (1996) and
Madau et al. (1996), among others.
We consider the contribution of all galaxies in our sampled volume
to the star-formation rate per unit comoving volume. The time-dependent
star-formation rate per any single galaxy is given by that of the  
best-fitting model
(i.e. Model 2 in the vast majority of galaxies, Model 1 for a 
couple of them), and by the formation redshift
$z_F$, everything scaled according to the best-fit baryonic mass $M$.

The contribution of each galaxy to the global comoving SF rate has been
estimated by dividing the time-dependent SF by the maximum
comoving volume $V_{max}$ within which the object would still be visible
above the sample flux limit.
We defer to Lilly et al. (1995, 1996) for a detailed treatment of how 
$V_{max}$ is computed.

A further correction factor to apply to the comoving SF rate takes into 
account the portion of the
luminosity function lost by the flux-limited sample within any
redshift interval (at any $z$ only galaxies brighter than K=20.1
can be detected). The correction is simply computed as the 
ratio of the total luminosity density to the luminosity-weighted integral 
of the LF above the luminosity corresponding to the flux limit at that $z$. 
This correction for completeness is not
very large (of the order of a few tens percent), because our adopted
K-band LLF is relatively flat at the faint end (see Connolly et al., 1997,
for further details on this point).

The global rate of SF $\Psi (z)$, which is the mass in stars formed per
year and unit comoving volume (expressed in solar masses per year and 
per cubic Mpc), is then the summed contribution by all 
galaxies in our sample. Figure 15 reports different estimates of  
$\Psi (z)$ based on the two assumptions of $q_0=0.15$ and $q_0=0.5$.  
The latter, in particular, is compared with estimates
by Lilly et al. (1996), Madau et al. (1996) and Connolly et al. (1997),
while a similar match is not possible for our best guess solution
with $q_0=0.15$. 

The calculation of $V_{max}$ and of the completeness correction depend
on the assumed rate of luminosity evolution.
The shaded regions in Fig. 15 mark two boundaries estimated with two  
different evolutionary 
corrections for luminosity evolution, to provide an idea of
the uncertainties. 
The lower limit is computed from Model 1 (moderate evolution at $z<3.5$, no
z-cutoff). As such, it provides a strict lower boundary to $\Psi (z)$.
For the upper curve, $V_{max}$ is computed after Model 2, i.e. assuming
stronger luminosity evolution and a cutoff in the available volume at
$z\simeq 1.5$. This upper curve corresponds to our best-guess $\Psi (z)$.
Unfortunately, a precise evaluation of the uncertainties is not
possible at this stage.

Figures 16 and 17 transform the information on the global SF rate
$\Psi (z)$ of Fig. 15 into one on the actual mass density in stars 
synthesized within any redshift interval (in solar masses
per unit comoving volume and per redshift interval). 
This is simply the time integral of the function $\Psi [z]$, and better 
quantifies when the various stellar populations are actually formed.
Figs. 16 and 17 correspond to the usual two choices for $q_0$, with 
$V_{max}$ computed from Model 2.
In each figure this information is also differentiated into various 
galactic mass intervals.

As it is apparent, there is a tendency for low-mass galaxies to have 
a star-formation activity protracted to lower redshifts. Indeed, several 
moderate to low-mass galaxies show young ages at low redshifts (see Fig. 7).
However,  although we have been as careful as possible in correcting for all
various selection and incompleteness effects, we cannot be conclusive about this,
in particular because of the limited size of our sample. Deeper and richer
samples are needed to settle the issue.

The dashed lines in the various panels of Figs. 16 and 17 provide the
integral distributions (reported on a linear vertical scale ranging from 0 to
100\% in each panel)
of the stellar mass synthesized as a function of $z$. For example, for $q_0=0.15$,
the dashed line in the top panel of Fig. 16 shows that in massive 
galaxies the SF is already over at $z=1$, while the third panel 
indicates that, for $M<5\ 10^{10}\ M_\odot$, 40\% of the stars are generated 
at $z<1$.

It is evident from Figs. 15 to 17 that the two cases considered, of an
Einstein-de Sitter or an open universe, entail somehow different
solutions for the evolutionary star-formation rate. In the closure world model
with $\Lambda=0$
there is a limited amount of time between the redshift of the observation
(typically $z\sim 1$) and the Big-Bang: this implies relatively higher
values of $z_F$ and a function $\Psi (z)$ keeping flat to the highest $z$.
An open universe, instead, with more cosmic time available, entails typically
lower values of $z_F$ and a $\Psi (z)$ peaked at slightly lower redshifts.

This reflects on the mass fractions of stars generated at the various
redshifts. For $q_0=0.15$, 80\% of the mass in stars is formed between
z=1 and z=3, while the remaining 20\% is made at higher or lower $z$.
 For $q_0=0.5$, the same 80\% of stars are formed between z=1.2 and z=4.2.
The median redshifts in the stellar formation process turn out to be
close to z=1.8 and z=2.5 in the two cases respectively.

The SF rate $\Psi (z)$ for our early-type field galaxy population
is compared in Fig. 15 with the average rate
(shaded horizontal regions) expected for early-type galaxies in rich clusters,
assumed that these synthesize between z=5 and 1.5 all metals observed 
in the intra-cluster medium (Mushotzky \& Loewenstein 1997). 
The comparison is mediated by a crucial
assumption we made on the stellar IMF to have a Salpeter's form. 
An IMF more  weighted in favor of massive stars would decrease the 
SF requirement for cluster ellipticals. 
There is in any case an indication that field ellipticals
behave as less efficient metal producers than the rich cluster counterparts.
It is quite conceivable that the large number of dynamical interactions
underwent by galaxies in the denser cluster environment might have triggered
a more intense SF during the early Gyrs of the cluster lifetime, with a
globally more efficient metal yield.

This evidence also matches a likely difference in the ages of cluster 
versus field ellipticals: while the properties of cluster galaxies imply
that the SF was completed already before z=2 (Stanford et al. 1998), 
typically 50\% of stars in the field are produced at more recent epochs.

Finally, it is interesting to compare in Figure 15 (lower panel) the 
indications on the SF history $\Psi (z)$, as inferred from direct 
inspection of high-redshift galaxies (see data points in the figure), 
with our predictions based on 
modelling of the stellar populations. The latter refer to only a sub-class 
of galaxies, those
classified as early-types. Yet, this prediction already exceeds the global
amount inferred by high-z "drop-outs".
There is a suggestion, in particular, that for $z>1.5$ the direct estimate
based on optical-UV measurements misses some light, if a sub-population alone
already exceeds the global estimated amount. This may be taken as an evidence
that dust indeed played a role in at least partly obscuring the young
stars during the active phase of star formation. If moderately 
true for the average
gas-rich spiral, this might have been particularly relevant during the
starburst events bringing to the formation of early-type galaxies.

\section{CONCLUSIONS AND PERSPECTIVES}

Early-type galaxies are commonly believed to contain among the oldest stellar 
systems in the universe. Analyses of galaxies in rich clusters,
which found weakly evolving spectra and non-evolving dynamical conditions
up to z=1 and slightly above, have confirmed this intuition.

In spite of a morphological similarity which may be taken as indicative of a 
common origin, no conclusive results were achieved so far about the early-type 
galaxies in the field. Our contribution here has been devoted to exploit
a near-IR sample 
of 35 distant early-types in the HDF with $K<20.15$, with optimal 
morphological information and spectro-photometric coverage, to study
properties of distant early-type galaxies outside rich clusters. 
Near-IR observations provide a view of galaxy evolution minimally biased
by the effects of the evolutionary K-correction, dust extinction and
changes in the $M/L$ ratio due to the aging of stellar populations. 
The basic conclusions of our analysis may be summarized as follows.

\begin{itemize}

\item 
The broad-band spectral energy distributions of the sample galaxies, 
together with the assumption of a Salpeter's initial mass function
within $M_l=0.15$ and $M_u=100\ M_\odot$, allow us to date their dominant 
stellar populations. The majority of
bright early-type galaxies in this field are found at redshifts $z
\lesssim 1.3$ to display colors indicative of a fairly wide range of
ages (typically 1.5 to 3 Gyrs).
This evidence adds to that of a substantially broad distribution of the
formation redshifts $z_F$ from galaxy to galaxy in the sample. So, there is
no coeval event of star-formation for early-types in this field, 
probably at variance with what happens for cluster galaxies.
There is a tendency for lower-mass systems to be typically younger than the
massive counterparts, but a firm conclusion has to wait for deeper and richer 
samples.
A spread in the ages and a somewhat protracted SF activity in field, as 
opposed to cluster, ellipticals may be consistent with their observed
narrow-band spectral indices ($H_\beta$, $Mg2$, etc.; see
Bressan, Chiosi \& Tantalo, 1996).

A protracted SF activity in field ellipticals may also bear some relationship
with their observed
preferentially "disky" morphology (Shioya and Taniguchi, 1993).
Unfortunately, our sample is too faint to allow testing this. Of the 15
galaxies (out of 35) bright enough to have the $c_4$ 
morphological parameter measured, 11 show rather clearly a "disky" morphology, 
and the other 4 a "boxy" shape.

\item
The basic uncertainty when dating stellar populations -- the metallicity of 
stars -- has been the subject of a careful discussion. Given the relatively 
blue colors of the detected galaxies at the typical redshift of one, 
the age uncertainty cannot be very large, as it would when considering
old populations in local galaxies. Our result should then be fairly
robust.

\item
Because of the different cosmological timescales, the redshift-dependent
star-formation history inferred from these data depends to some extent on the
assumed value for the cosmological deceleration parameter. 
We find that the major episodes of star-formation
building up 80\% of the mass in stars for typical $M^{\star}$ galaxies 
have taken place during the wide
redshift interval between $z=1$ and $z=4$ for $q_0$=0.5, which becomes $z=1$ 
to $z=3$ for $q_0$=0.15.
Lower-mass ($M<5\ 10^{10}\ M_{\odot}$) systems tend to have their bulk of SF 
protracted to lower redshifts (down to almost the present time).

\item
The previous items dealt with the ages of the dominant stellar populations
in distant field galaxies. The evolution history for the dynamical 
assembly of these stars into a galactic body might
have been entirely different. However, our estimated galactic
masses, for a Salpeter IMF, are found in the range from a few $\sim 10^{9}\
M_\odot$ to a few $10^{11}\ M_\odot$ already at $z\simeq 1$.  So the
massive end of the E/S0 population appears to be mostly in place by that 
cosmic epoch, with space densities, masses and luminosities consistent with
those of the local field population. 
It is to be investigated how this result compares with published findings
(see e.g. Kauffmann \& Charlot 1998)
of a strong decrease of the comoving mass density of early-type
galaxies already by $z\simeq 1$. We suggest that a question to keep under
scrutiny in these analyses concerns the color classification, as we find,
after a detailed morphological and photometric analysis, that these objects 
usually display blue young populations mixed with old red stars.

\item
The present sample is characterized by a remarkable absence of objects at 
$z>1.3$, which should be 
detectable during the luminous star-formation phase expected to
happen at these redshifts. This conclusion seems robust:
if optical spectroscopy has difficulties to enter this redshift
domain, our analysis, using broad-band galaxy SEDs when spectroscopic
redshifts are not available, has no obvious biases.
The uncertainty in the photometric estimate of $z$ for this kind
of spectra is small. We discuss solutions for this sudden disappearance
in terms of: 
{\it a)} merging events, triggering the SF, which imply strongly perturbed 
morphologies and which may prevent selecting them by our
morphological classification filter, and {\it b)} a dust-polluted
ISM obscuring the (either continuous or episodic) events of star-formation, 
after which gas consumption (or a galactic wind) cleans up the galaxy.
We conclude that the likely solution is a combination thereof, 
i.e. a set of dust-enshrouded merging-driven starbursts occurred during
the first few Gyrs of the galaxy's lifetime.

\item
A comparison between the observed SF rates with the level needed to synthesize
the metals in the ICM indicates that field ellipticals could have been 
slightly less efficient
metal producers than the cluster galaxies. This difference adds to the
one in the ages and age spread.

\item
The comparison in Fig. 15b of our estimated SF rate $\Psi (z)$ 
at $z>1.5$ with those inferred
from the optical colors of high-redshift galaxies (via the Lyman "drop-out" 
or other photometric techniques) provides a {\it direct indication that a 
fraction of light
emitted during the starburst episodes (in particular those concerning the
formation of early-type galaxies) has been lost, probably obscured by dust}. 
The recently detected IR/sub-millimetric
extragalactic background (Puget et al. 1996; Hauser et al. 
1997) may be a trace
of this phenomenon. Our results on the cosmological SF rate reported 
in Fig. 15b (note that our best-guess coincides with the upper limit
of the shaded region) are very close to the level predicted by Burigana
et al. (1997) to reproduce the spectral intensity of the cosmic 
infrared background.

It will obviously be essential to have a direct test of such a 
dust-extinguished SF, but this will not be easy until powerful
dedicated instrumentation will be available. 
The difficulty is illustrated in Figure 18, 
showing possible spectra corresponding to various ages of
a massive ($M=3\ 10^{11}\ M_\odot$) galaxy at $z=2$, according to Model 2, 
during the starburst and post-starburst phases. 
As indicated by the figure, various planned missions
(in launch-time order, SIRTF, ESA's FIRST,
NASA's NGST, and some ground-based observatories in
exceptionally dry sites, like the South Pole) could discover 
intense activity of star-formation at $z=1.5$ to 3 or 4.
Also interesting tests of these ideas could be soon achieved with SCUBA 
on JCMT, and perhaps with ISO.
In any case, observations at long wavelengths will be needed to completely
characterize the early evolutionary phases of (spheroidal) galaxies.

\item
While our main conclusions are moderately dependent on the
assumed value of $q_0$, an open universe or one with non-zero $\Lambda$ are 
favored in our analysis by 
the match of the K-band local luminosity functions with the observed 
numbers of faint distant galaxies.  

\item
There are two main sources of uncertainty in our analysis. 
The redshift distribution in Figure 12 (and in particular the
pronounced peak at $z=1.1$) may be somehow affected
by the possible presence of a background cluster or group at $z\sim 1$.
A few to several of the galaxies with photometric redshifts in the interval 
$1<z<1.2$ have a non-random distribution in the HDF, being clustered
in one region (at RA$\simeq$87.98, DEC$\simeq$62.218). 
Until we will not have a complete
spectroscopic identification of most of the galaxies with precisely
measured redshifts, it will be impossible to check this. However the
effect should be limited to a small fraction of the sample galaxies
and should not drastically influence our statistics.
The second basic uncertainty is due to the photometric estimate of the
redshift for half of the sample. In this case too, however, this is not likely
to affect our main conclusions. In any case, these uncertainties will be 
reduced soon by 
new observations in the Southern HDF and deep spectroscopic surveys from 
ground with large telescopes.

\end{itemize}

%
%

\acknowledgments
This work has been supported by the European Community with the TMR Grants
ERBFMRX-CT96-0086 and ERBFMRX-CT96-0068.
We also acknowledge partial funding by the Italian Space Agency under 
Contract ASI-ARS-9689. We are indebted to an anonymous referee for very helpful 
comments on the paper.


\newpage





\epsfxsize=14cm
\begin{figure}[!Ht]
\vspace*{-10pt}
\hspace*{0pt}
\epsffile{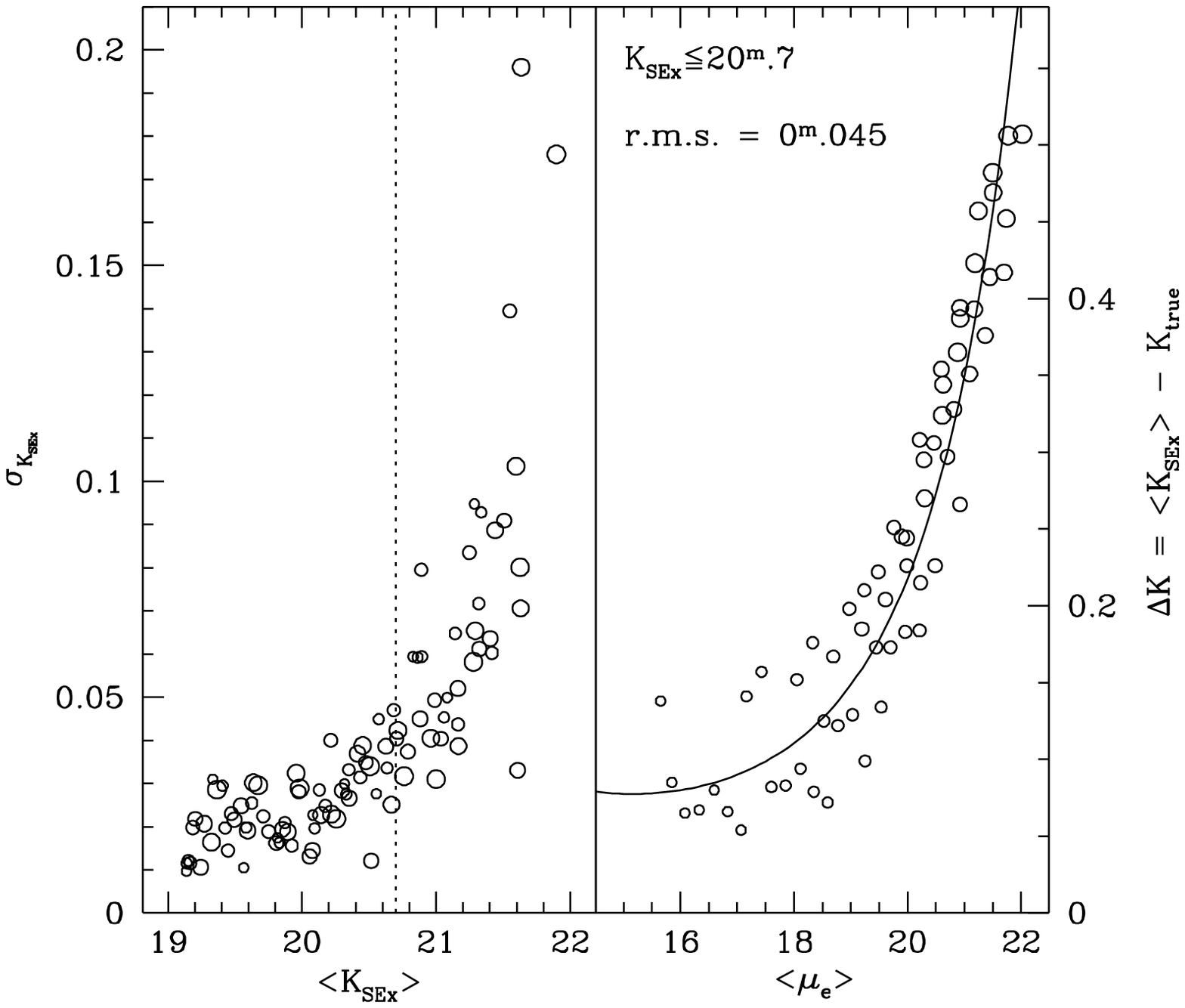}
\vspace*{-10pt}
\caption {Panel (a): standard deviation of the SExtractor magnitude 
estimates as a function of the average magnitude for galaxies in the 
simulated images: the standard deviation is very small for 
$<K_{SEx}>\leq 20^m.7$.
Panel (b): difference between true flux and the SExtractor flux as a 
function of the average surface brightness for the subsample of 
simulated galaxies with $<K_{SEx}>\lesssim 20^m.7$: it shows that
only very low-surface brightness objects ($<\mu_e^K(SEx)>\geq 22$) could be
missed by the IRIM $K-band$ image.}
\label{fig1}
\end{figure}

\clearpage\newpage

\epsfxsize=14cm
\begin{figure}[!Ht]
\vspace*{-10pt}
\hspace*{0pt}
\epsffile{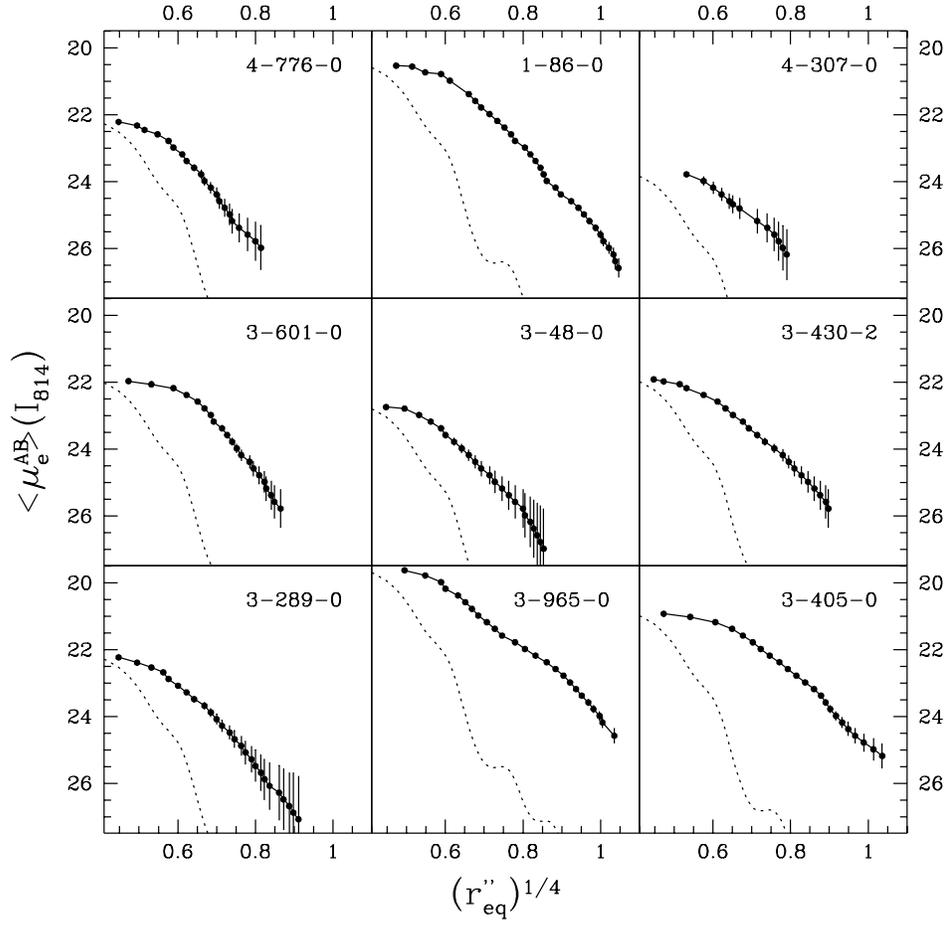}
\vspace*{-10pt}
\caption{Luminosity profiles in the $I_{814}$ for nine galaxies in the 
sample. The dotted kine is a fit to the PSF.
}
\label{fig2}
\end{figure}

\clearpage\newpage

\epsfxsize=14cm
\begin{figure}[!Ht]
\vspace*{-10pt}
\hspace*{0pt}
\epsffile{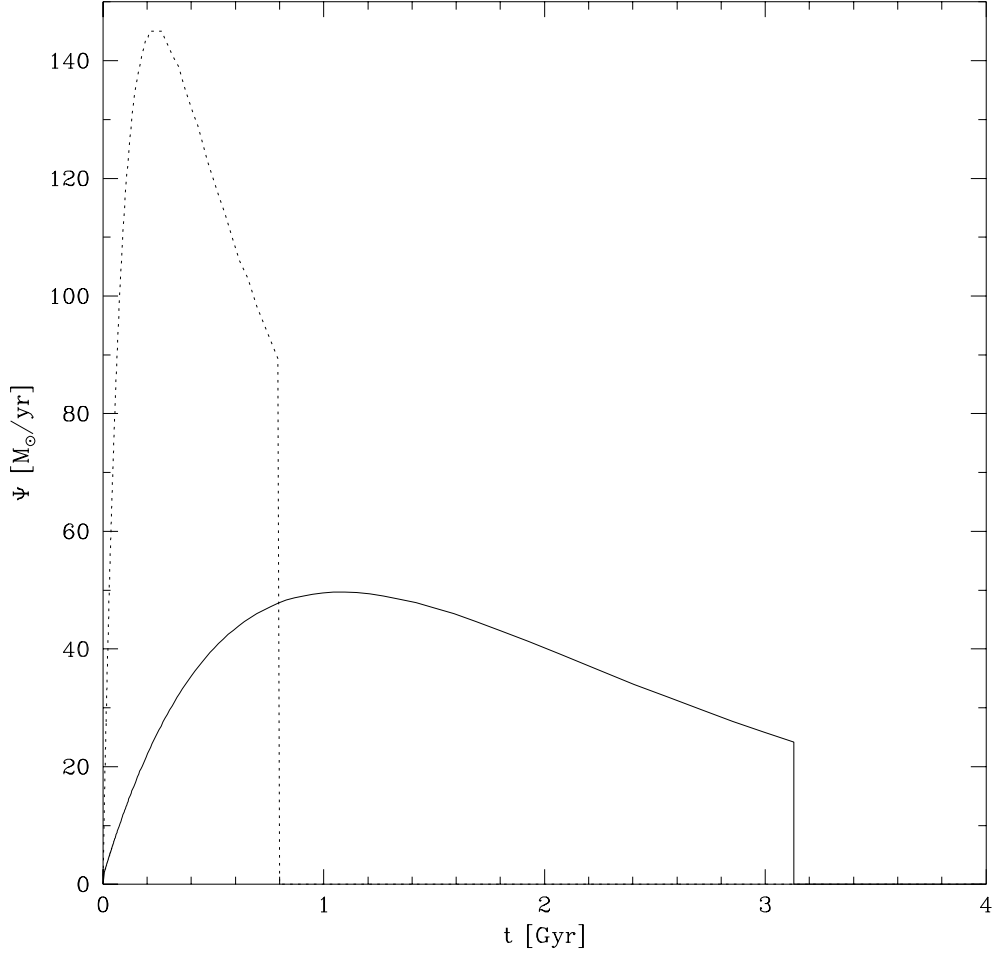}
\vspace*{-10pt}
\caption{The star-formation rate as a function of the galactic
time for our two reference models (dotted line: Model 1;
continuous line: Model 2). The fall-off at t=0.8 and 3.15 Gyrs corresponds
to the onset of the galactic wind. The normalization is for a galactic
baryonic mass of $M=10^{11}\ M_\odot$.
}
\label{fig3}
\end{figure}

\clearpage\newpage

\begin{figure}[!Ht]
\vspace*{-10pt}
\hspace*{0pt}
\epsffile{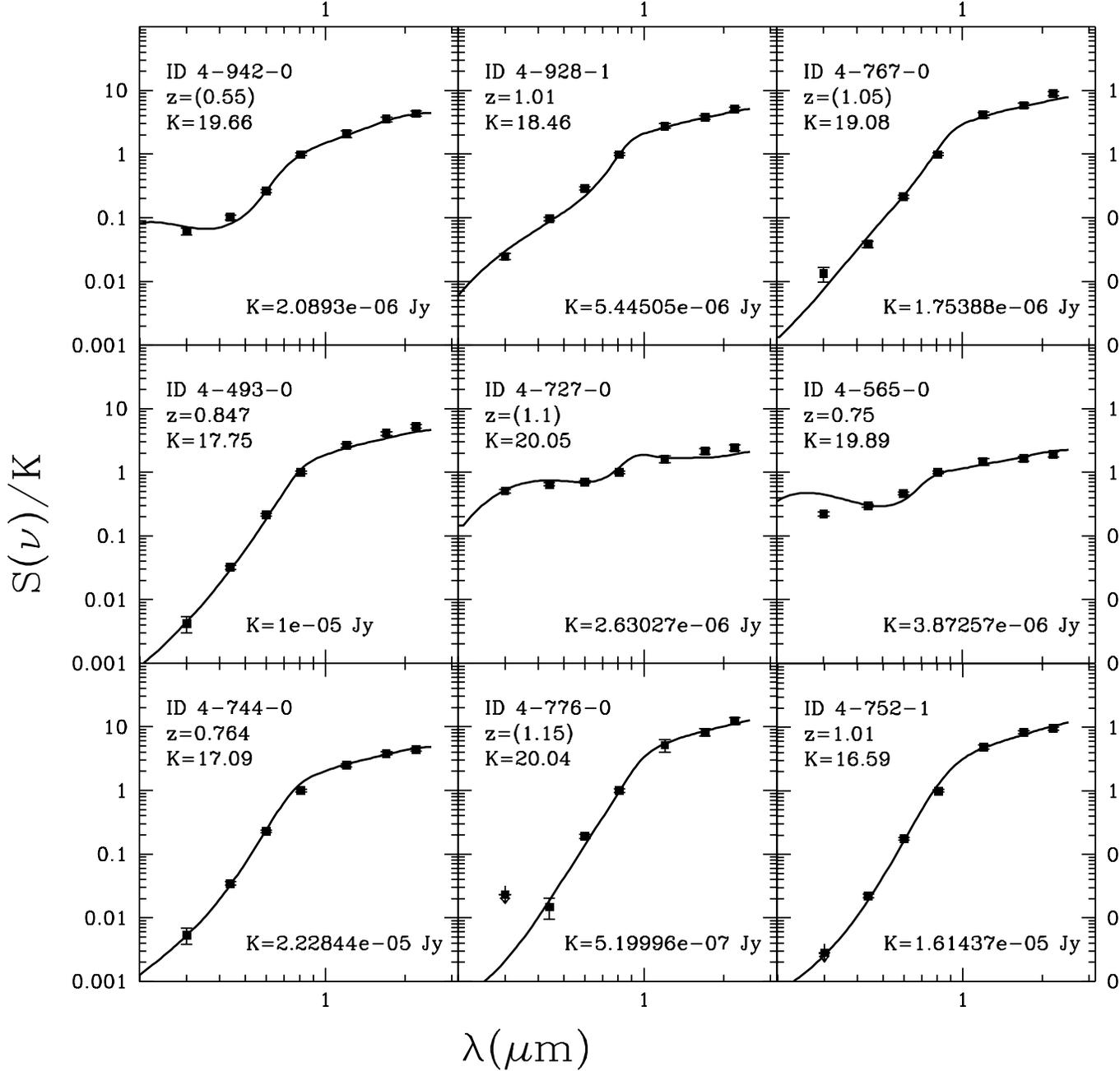}
\vspace*{-10pt}
\caption{Observed broad-band spectra for all sample galaxies, fitted with
Model 2 for $q_0=0.15$. For some objects we add to Model 2 a residual SF 
parametrized by $\Psi_0$ in  Table 2.
For each object we provide the name, redshift (in parenthesis if based on
photometric fits), K magnitude, and the normalization factor of the y-axis.
}
\label{fig4}
\end{figure}

\clearpage\newpage

\epsfxsize=14cm
\begin{figure}[!Ht]
\epsffile{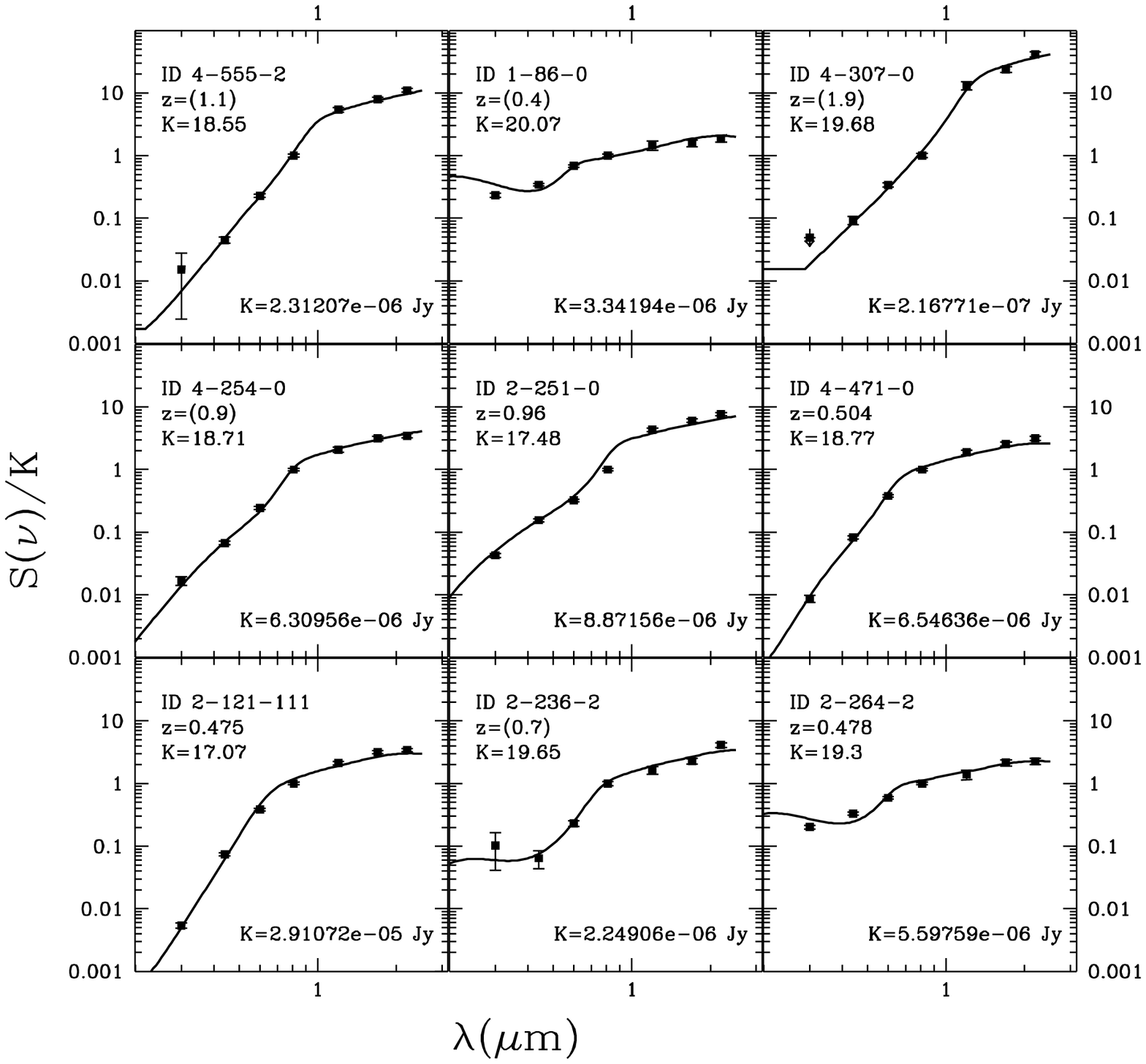}
\centerline{Fig. 4b}
\end{figure}

\clearpage\newpage

\epsfxsize=14cm
\begin{figure}[!Ht]
\epsffile{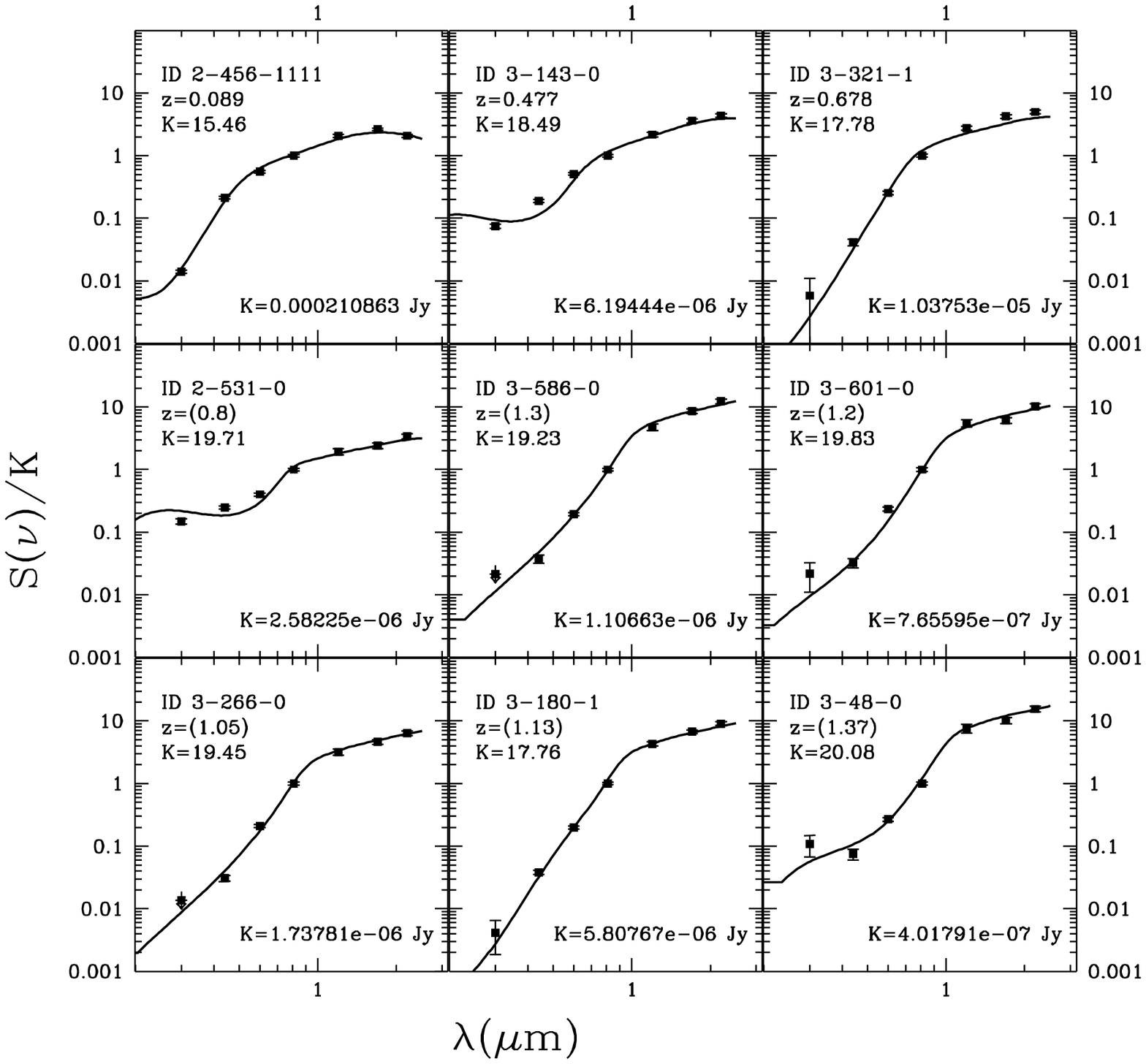}
\centerline{Fig. 4c}
\end{figure}

\clearpage\newpage

\epsfxsize=14cm
\begin{figure}[!Ht]
\epsffile{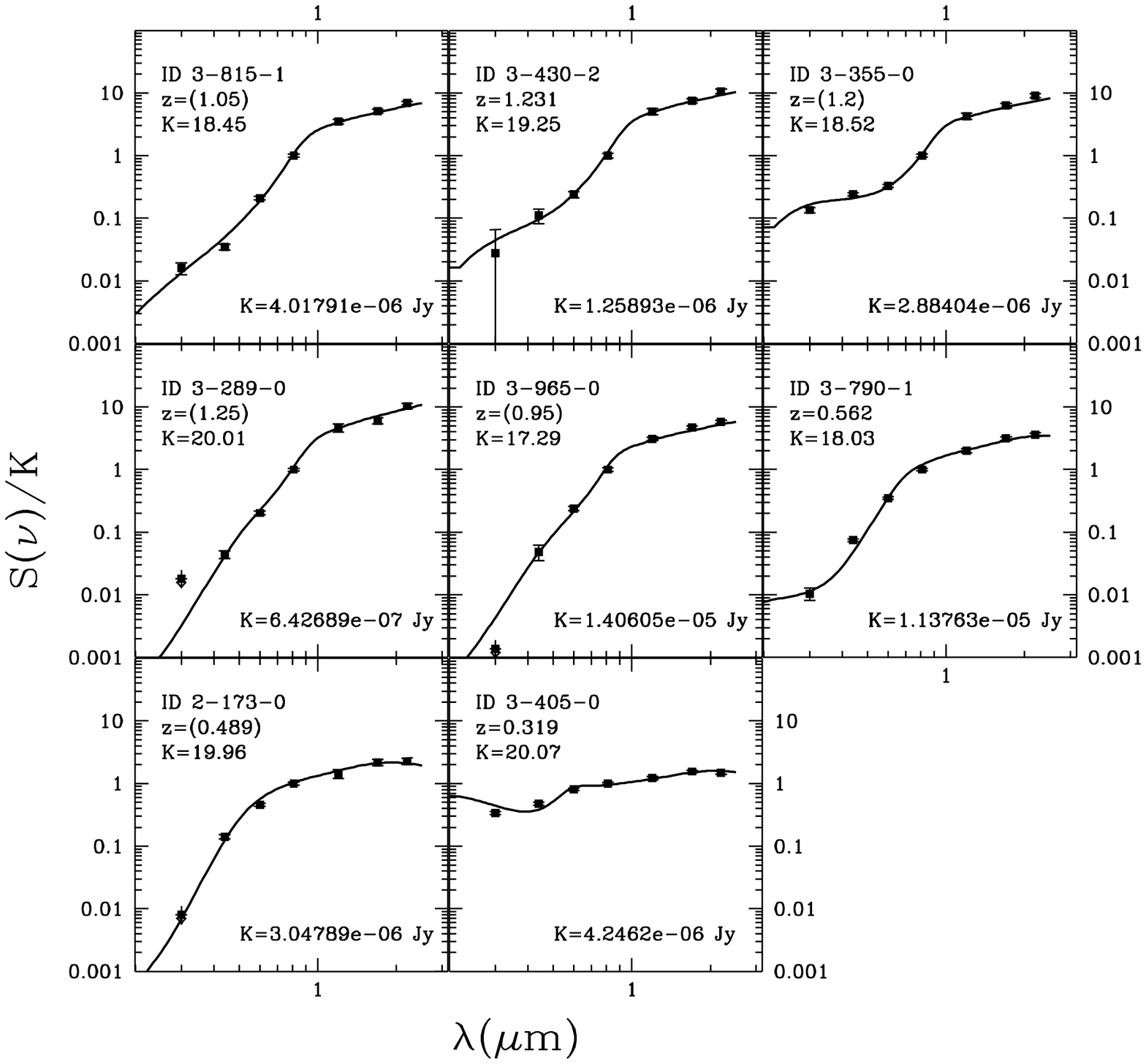}
\centerline{Fig. 4d}
\end{figure}

\clearpage\newpage

\epsfxsize=14cm
\begin{figure}[!Ht]
\vspace*{-10pt}
\hspace*{0pt}
\epsffile{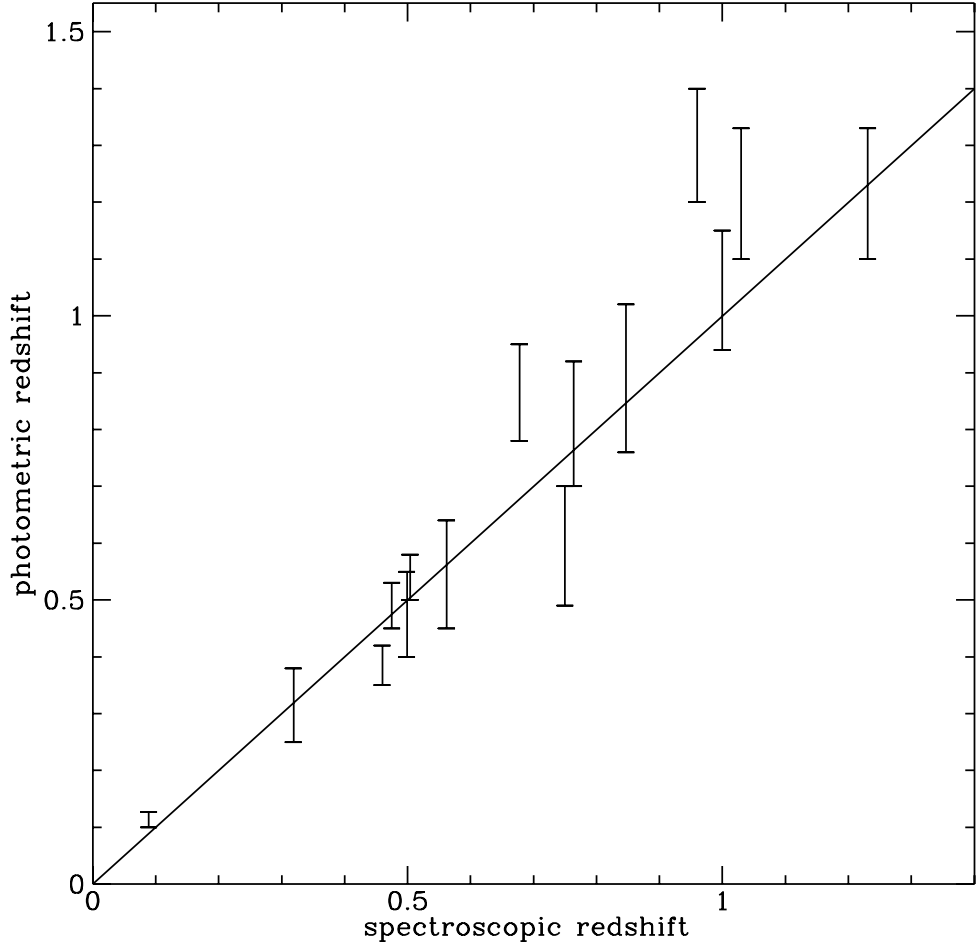}
\vspace*{-10pt}
\caption{Comparison of photometric redshifts, based on 7-band spectral data,
with spectroscopic redshifts. Errorbars are $\sim95\%$ intervals based on
$\chi^2$ fitting of Model 2. The discrepant object at $z=0.96$ (ID 2-251-0)
contains a nuclear non-thermal source, and it is the only galaxy in our sample 
with this feature.
}
\label{fig5}
\end{figure}

\clearpage\newpage

\epsfxsize=14cm
\begin{figure}[!Ht]
\vspace*{-10pt}
\hspace*{0pt}
\epsffile{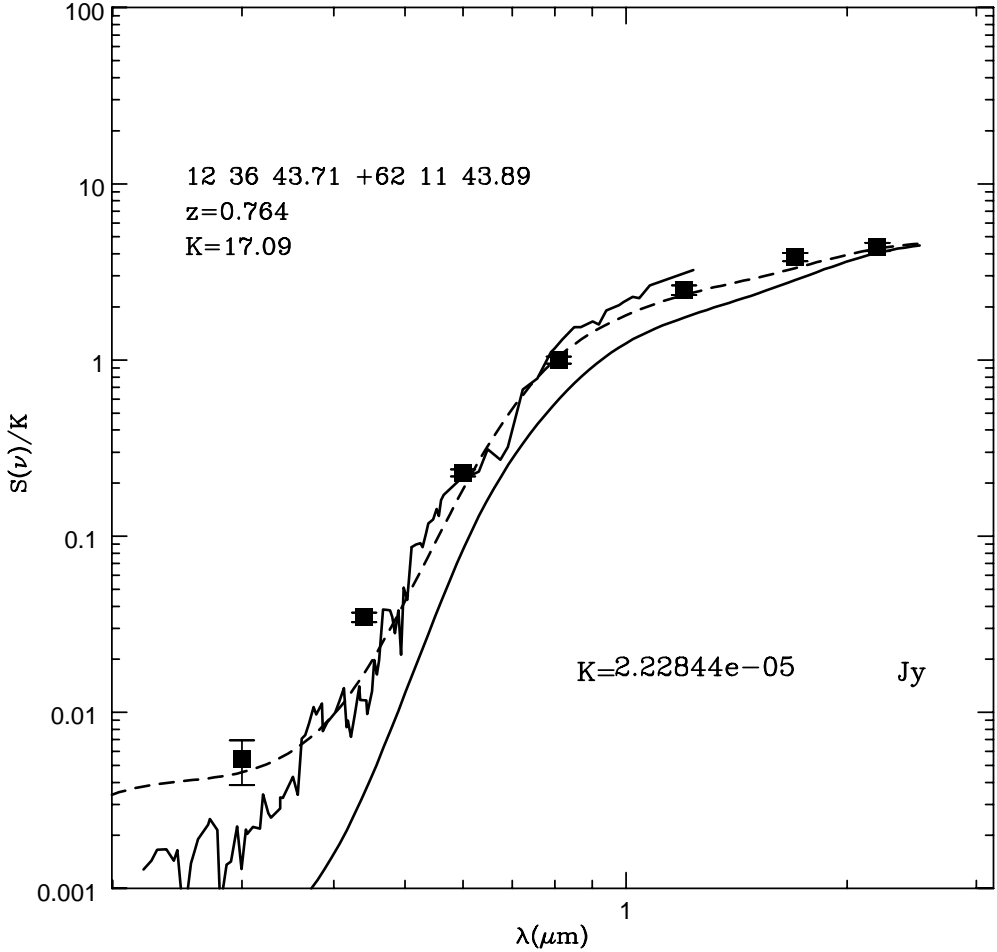}
\vspace*{-10pt}
\caption{Spectrum of a typical class [b] galaxy in the sample compared
with two synthetic spectra based on Model 1 (with $q_0=0.15$, see dashed line). 
Continuous line: 
case with $z_F=5$; dashed line: $z_F=3$. This is compared with the optical 
spectrum of the local early-type M32 (see text for more details).
}
\label{fig6}
\end{figure}

\clearpage\newpage

\epsfxsize=14cm
\begin{figure}[!Ht]
\vspace*{-10pt}
\hspace*{0pt}
\epsffile{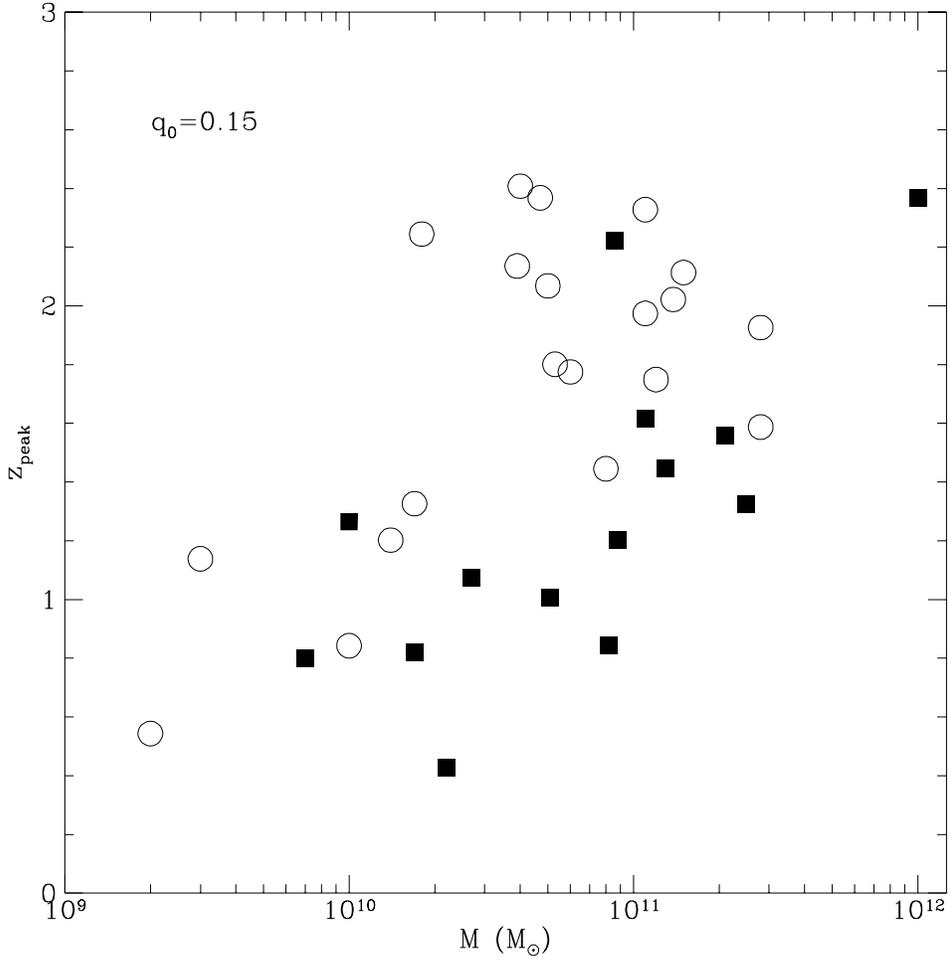}
\vspace*{-10pt}
\caption{Baryonic mass versus redshift of peak star-formation, according
to spectral best-fits based on Model 2. Panel (a): 
the solutions assume $q_0=0.15$; Panel (b): 
solutions based on $q_0=0.5$ (see Table 2 for the best-fit parameters). 
Filled squares
refer to objects with spectroscopic redshift, open circles to those
with photometric redshift.
}
\label{fig7a}
\end{figure}

\clearpage\newpage

\epsfxsize=14cm
\begin{figure}[!Ht]
\epsffile{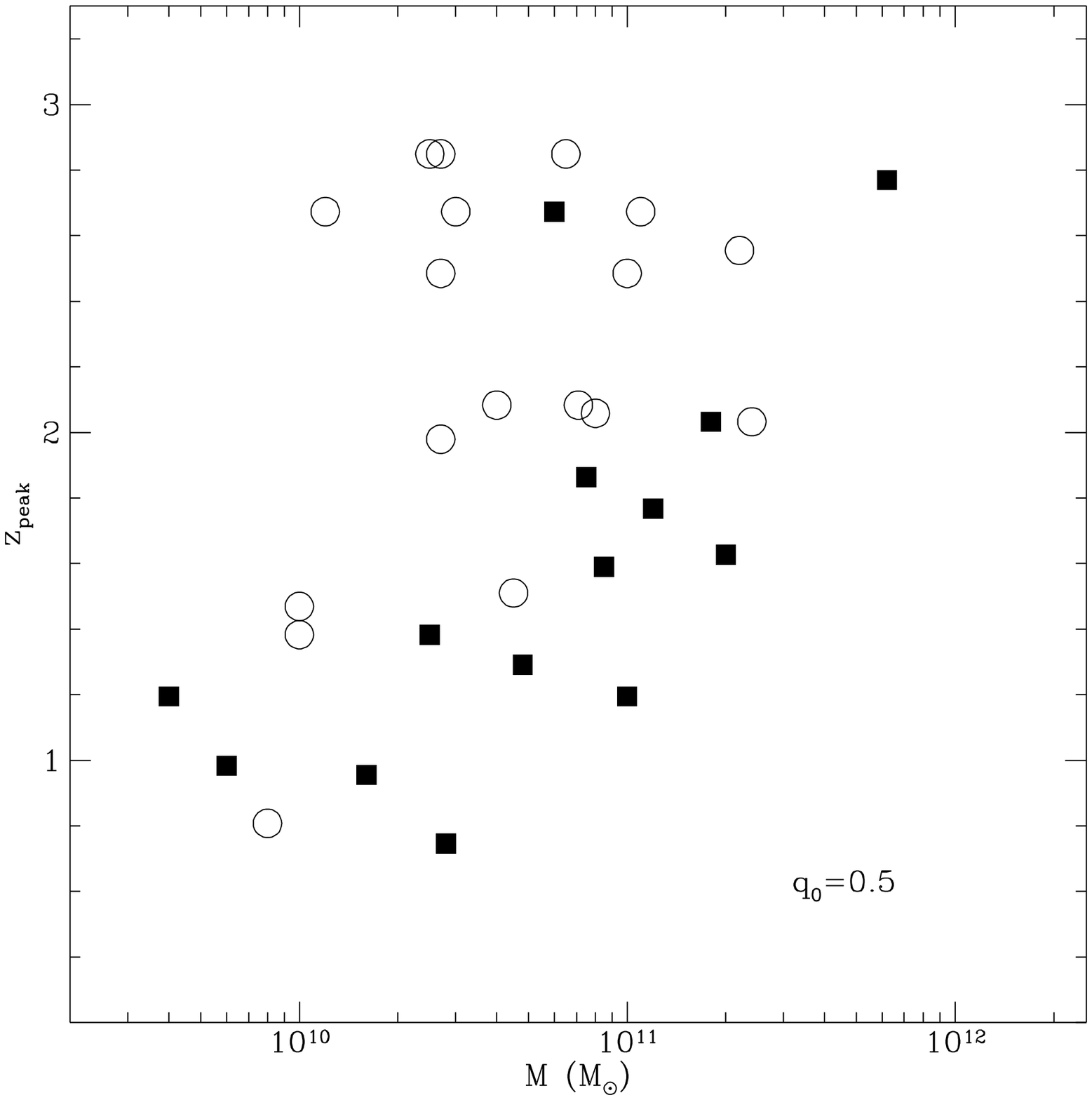}
\centerline{Fig. 7b}
\end{figure}

\clearpage\newpage

\epsfxsize=14cm
\begin{figure}[!Ht]
\vspace*{-10pt}
\hspace*{0pt}
\epsffile{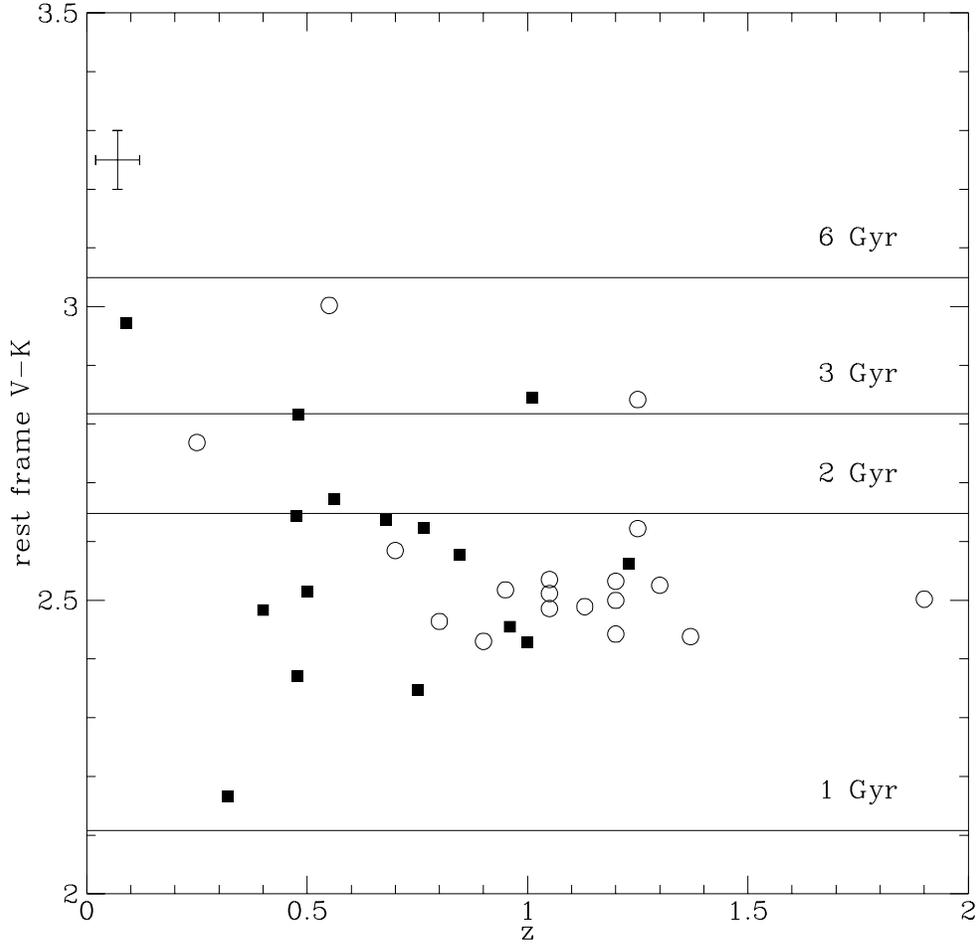}
\vspace*{-10pt}
\caption{Rest-frame V--K (panel [a]) and B--J (panel [b]) colors, 
compared with predicted values for single stellar populations with 
solar metallicity. The ages for the latter are indicated, as well as
the mean color of local galaxies. Meaning of the symbols as in Figure 7.}
\label{fig8a}
\end{figure}

\clearpage\newpage

\epsfxsize=14cm
\begin{figure}[!Ht]
\epsffile{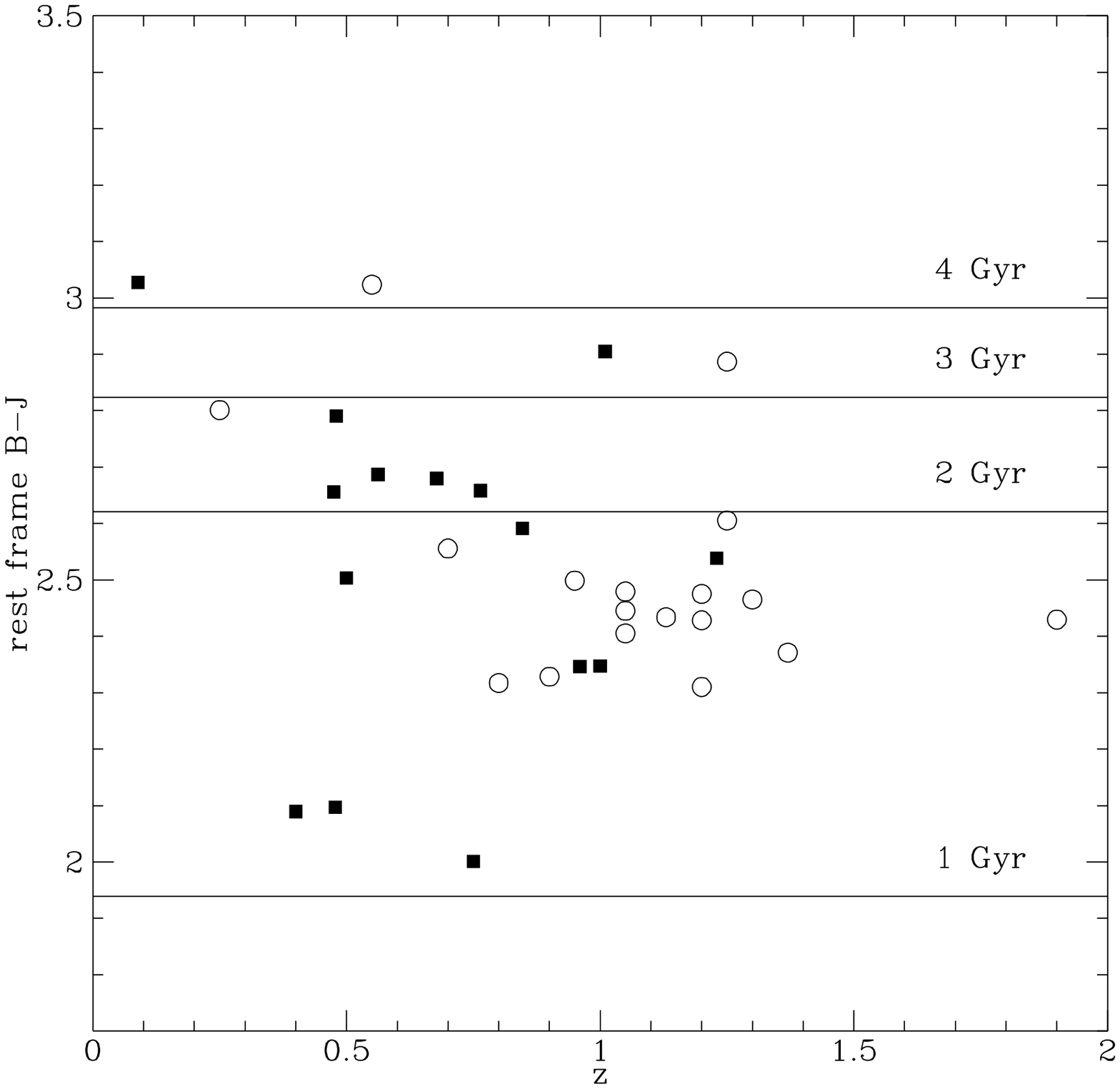}
\centerline{Fig. 8b}
\end{figure}

\clearpage\newpage

\epsfxsize=14cm
\begin{figure}[!Ht]
\vspace*{-10pt}
\hspace*{0pt}
\epsffile{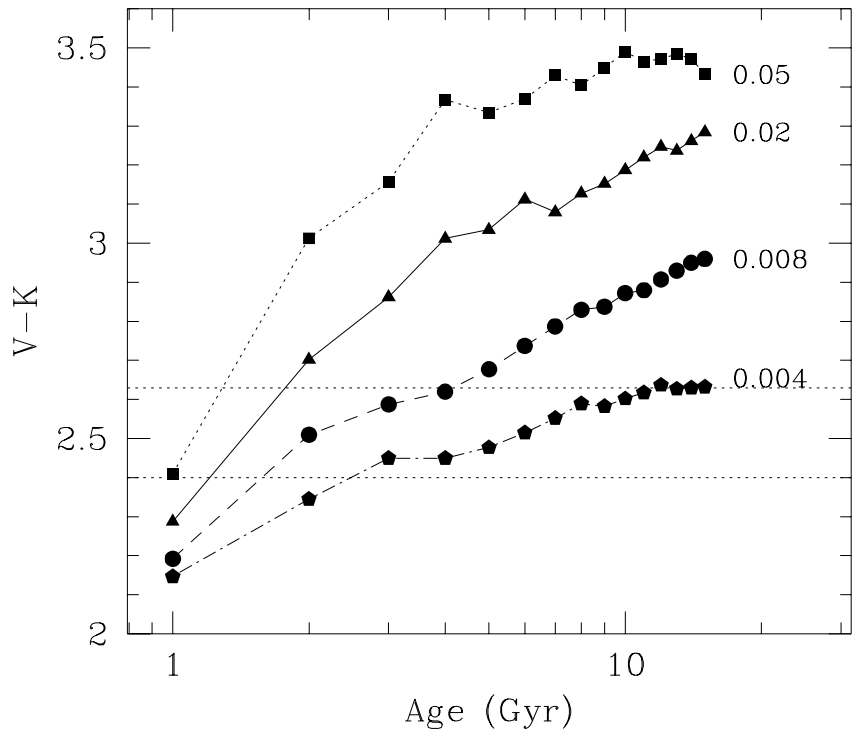}
\vspace*{-10pt}
\caption{Predicted color evolution of single stellar populations as a function 
of galactic age for different metallicities (indicated in the figure). 
Panel (a): V-K; Panel (b): B-J.
The dotted horizontal lines mark the boundaries of the observed distribution.
}
\label{fig9a}
\end{figure}

\clearpage\newpage

\epsfxsize=14cm
\begin{figure}[!Ht]
\epsffile{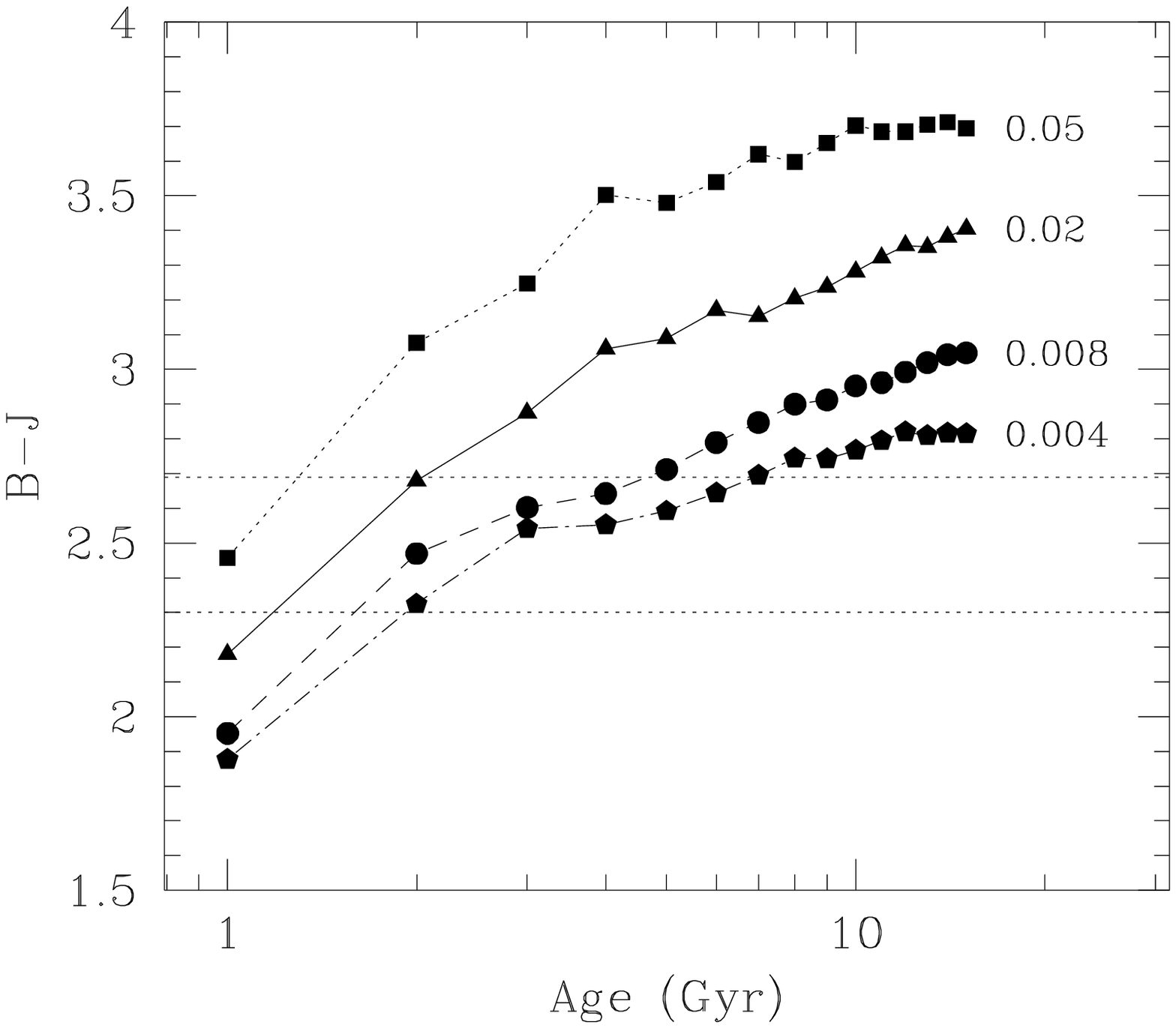}
\centerline{Fig. 9b}
\end{figure}

\clearpage\newpage

\epsfxsize=14cm
\begin{figure}[!Ht]
\vspace*{-10pt}
\hspace*{0pt}
\epsffile{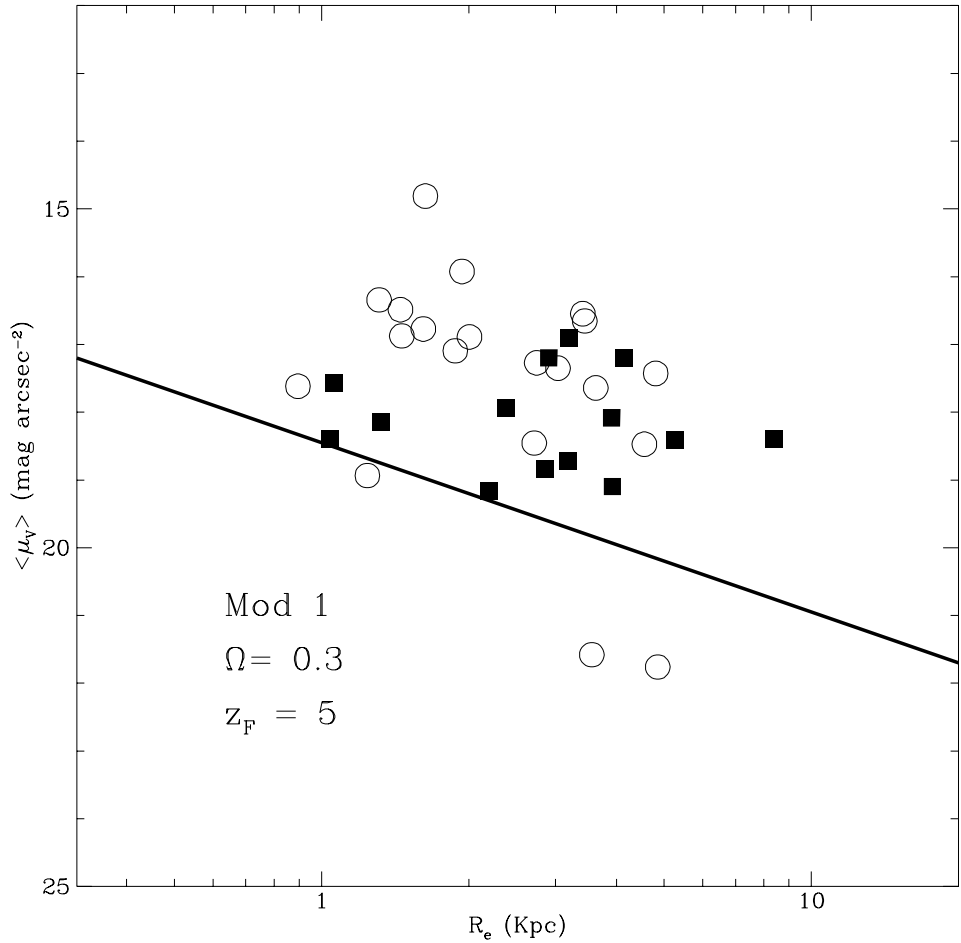}
\vspace*{-10pt}
\caption{The Kormendy relation in the V band, i.e. the average surface
brightness versus effective radius. The continuous line represents the local
relation. The data are obtained applying to the observed surface brightness
K- and evolutionary corrections according to Model 1
($q_0=0.15$, $z_F=5$). Meaning of the symbols as in Figure 7.
}
\label{fig10}
\end{figure}

\clearpage\newpage

\epsfxsize=14cm
\begin{figure}[!Ht]
\vspace*{-10pt}
\hspace*{0pt}
\epsffile{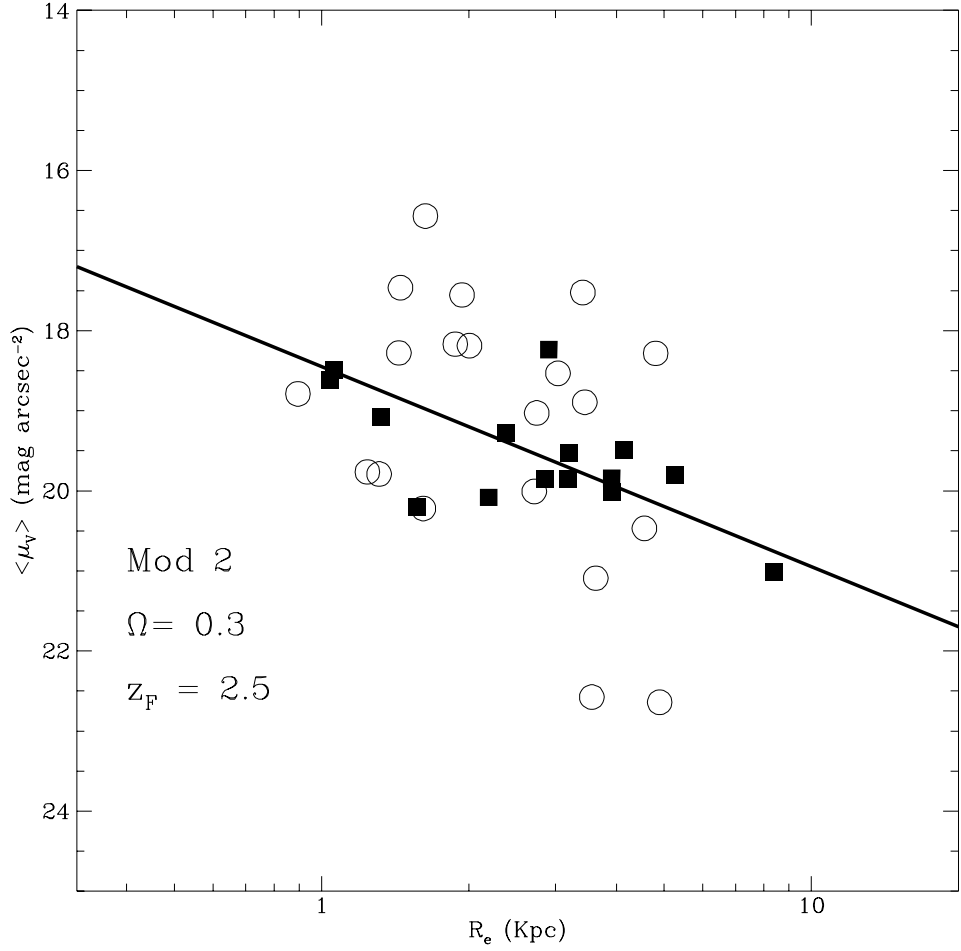}
\vspace*{-10pt}
\caption{Panel (a): same as the previous figure, but correcting the 
observed surface brightness according to Model 2
($q_0=0.15$, $z_F=2.5$). Panel (b): Kormendy relation in the K band, data
corrected as in panel (a). The mean locus of local galaxies comes from
Pahre et al. (1995).
}
\label{fig11}
\end{figure}

\clearpage\newpage

\epsfxsize=14cm
\begin{figure}[!Ht]
\epsffile{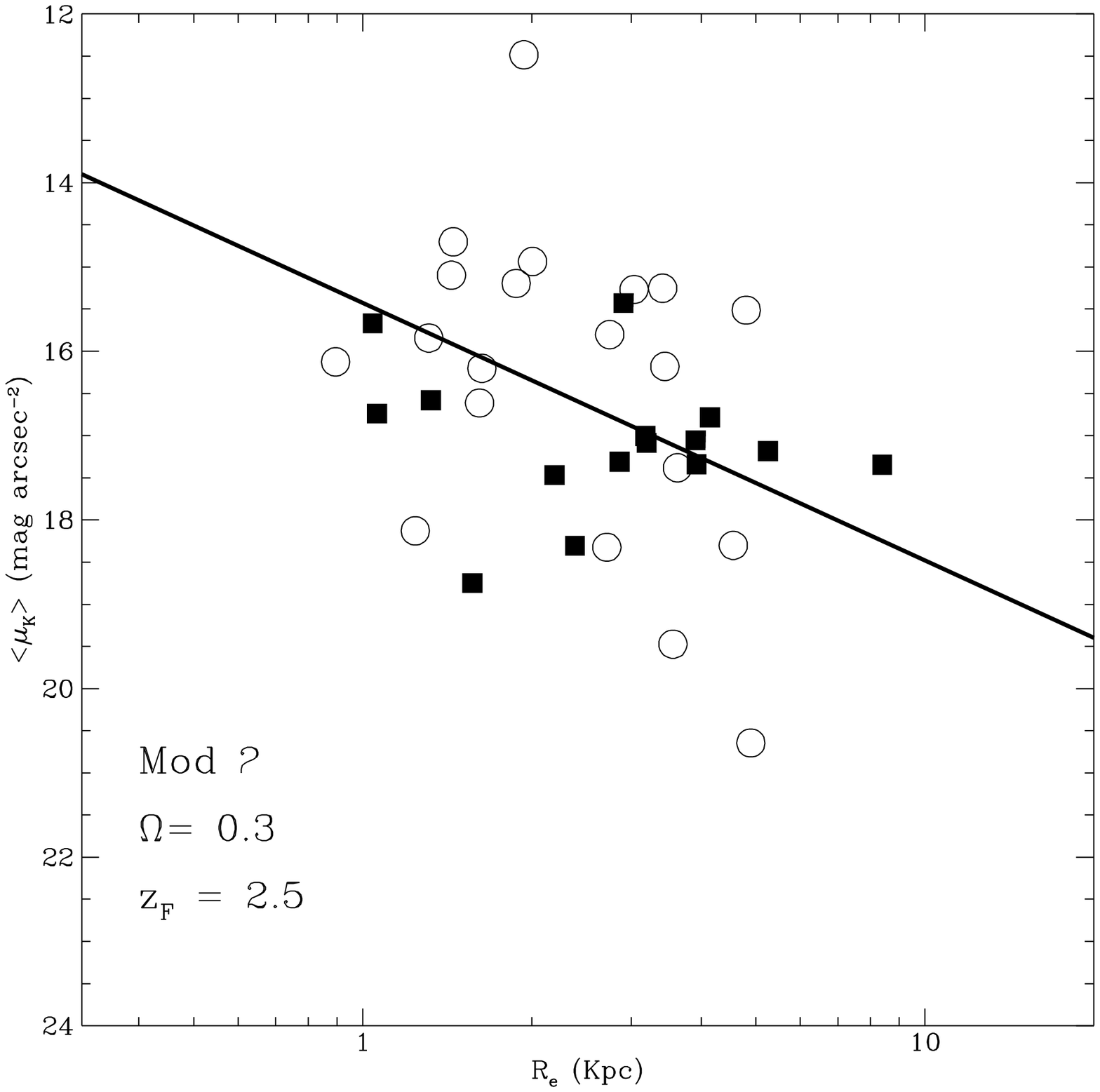}
\centerline{Fig. 11b}
\end{figure}

\clearpage\newpage

\epsfxsize=14cm
\begin{figure}[!Ht]
\vspace*{-10pt}
\hspace*{0pt}
\epsffile{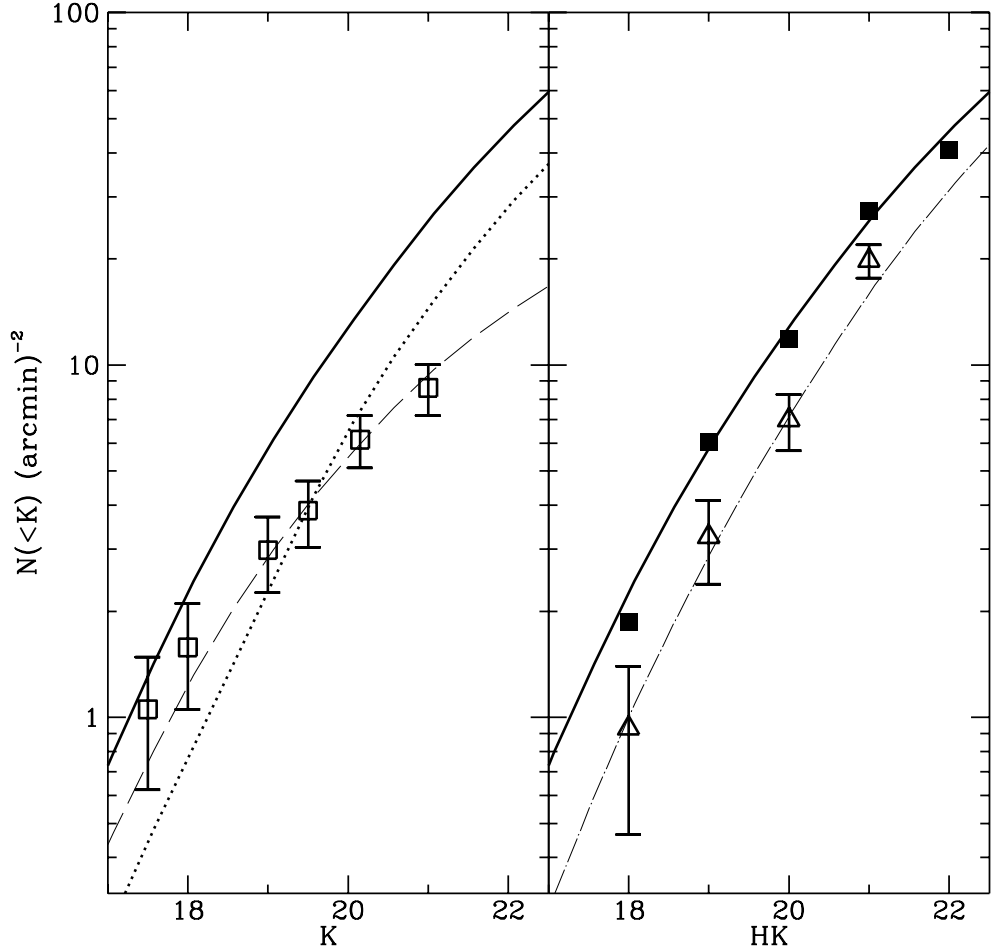}
\vspace*{-10pt}
\caption{ Counts as a function of the morphological type in the K and HK
bands. Open squares in panel (a) are the counts for E/S0 coming from
our sample. The datum at K=21 is based on the HDF Active Catalogue by
Cowie et al. (1996). These data are compared with the predicted counts
by to Model 1 with $q_0=0.15$, $z_F=4$ (dotted line), and with
counts by Model 2 with dust extinction during the SF phase (dashed line). 
A comparison is
made with the total predicted counts from Model 2 (thick continuous line).
In panel (b) we report counts in the H+K band for galaxies classified as 
spirals and irregulars
in the catalogue by Cowie et al. (open triangles), compared with the
total counts (filled squares). The dot-dashed line is based on a
moderately evolving model for late-type galaxies as discussed in the text.
}
\label{fig12}
\end{figure}

\clearpage\newpage

\epsfxsize=14cm
\begin{figure}[!Ht]
\vspace*{-10pt}
\hspace*{0pt}
\epsffile{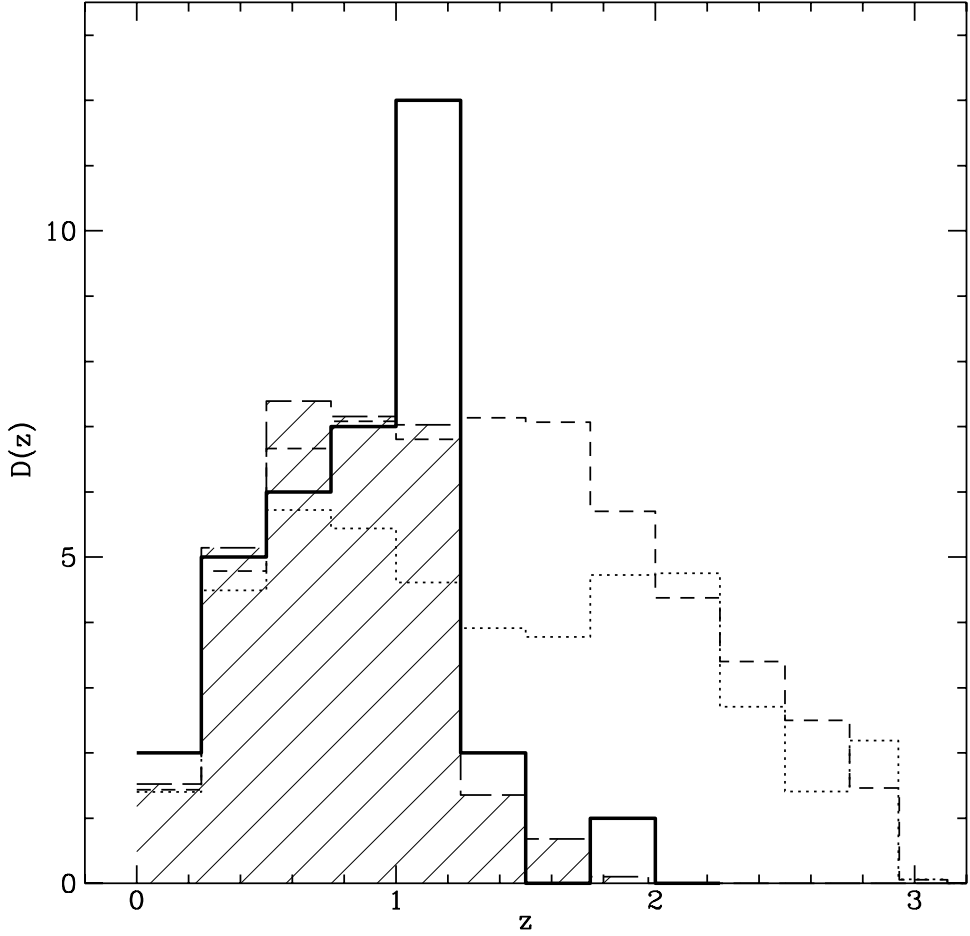}
\vspace*{-10pt}
\caption{ Redshift distributions. The continuous-line histogram is our observed 
distribution, for $K<20.15$, over an area of 5.7 square arcmin. 
The dotted line is the prediction of Model 1 ($q_0=0.15$, $z_F=4$). 
For Model 2 without dust extinction (short-dashed line) the misfit is even more severe.
Dashed line-shaded region: predicted distribution according to Model 2 
with a dusty SF phase.
}
\label{fig13}
\end{figure}

\clearpage\newpage

\epsfxsize=14cm
\begin{figure}[!Ht]
\vspace*{-10pt}
\hspace*{0pt}
\epsffile{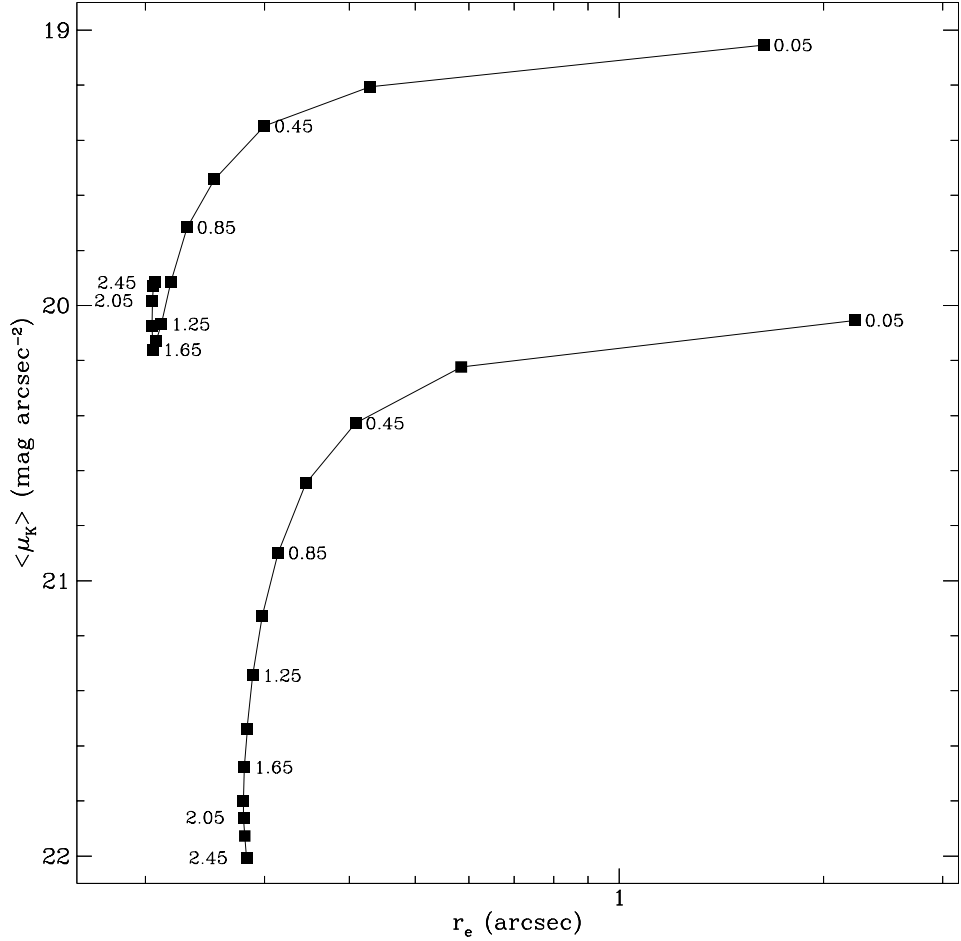}
\vspace*{-10pt}
\caption{Scaling of the surface brightness and effective radius as a function 
of redshift. Two paths are shown: the top one corresponds to a typical galaxy
in our sample
allowed to evolve following Model 2. The bottom line corresponds to 
the lowest surface
brightness object in the sample evolving according to Model 1
($q_0=0.15$, $z_F=5$). The stronger decrease with redshift of the 
average surface brightness in this second case is due to a strong K-correction
and almost no luminosity evolution implied by this model.
The latter represents a very conservative evolution patter, yet the
galaxy would remain detectable up to $z\simeq 2.5$ above the limiting
brightness indicated by the simulations (see Sect. 2.1). It is then 
unlikely that the surface brightness limitation could affect
the selection of E/S0 galaxies above z=1.3.
}
\label{fig14}
\end{figure}

\clearpage\newpage

\epsfxsize=14cm
\begin{figure}[!Ht]
\vspace*{-10pt}
\hspace*{0pt}
\epsffile{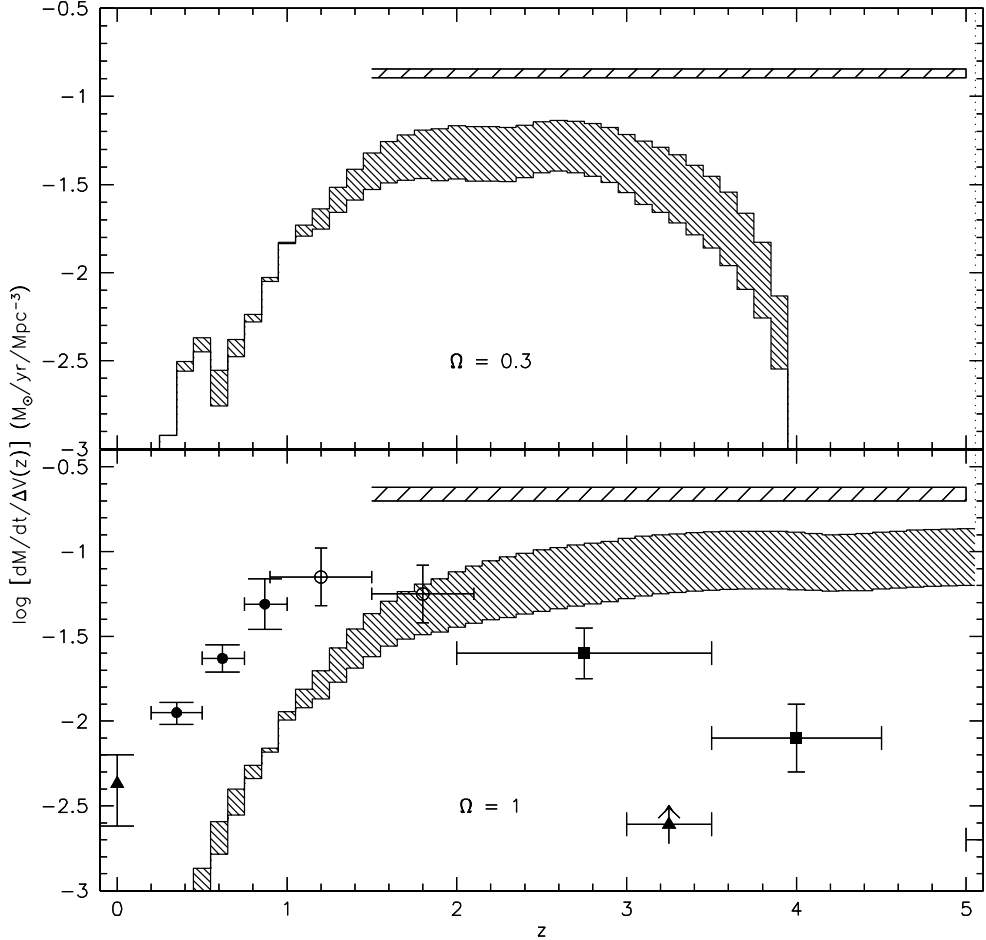}
\vspace*{-10pt}
\caption{
The star-formation history per unit comoving volume and unit time
$\Psi (z)$ of early-type galaxies, as inferred from 
our analysis of the K-HDF sample. Panel (a): the thick shaded region
is bounded by two solutions (for $q_0=0.15$) corresponding to two
different estimates of the maximum volume $V_{max}$ available to each
source. The lower curve is a limit based on a very conservative
estimate of $V_{max}$ assuming no redshift cutoff for E/S0's. The upper curve 
corresponds to our best guess for $\Psi (z)$ computed assuming a cutoff
in $V_{max}$ at $z\sim 1.5$ (as implied by our statistical analysis). 
The shaded horizontal region marks the universal rate of star formation
(for our adopted Salpeter IMF) estimated by Mushotzky \& Loewenstein
(1997), assuming that all field E/S0 galaxies behave as the cluster
objects in terms of the efficiency of metal production. 
Panel (b) is the same as panel (a) for our best-fit solutions with 
$q_0=0.15$. Here a comparison is possible with independent evaluations
of $\Psi(z)$ by Lilly et al. (1996), Madau et al. (1996) and Connolly et al.
}
\label{fig15}
\end{figure}

\clearpage\newpage

\epsfxsize=14cm
\begin{figure}[!Ht]
\vspace*{-10pt}
\hspace*{0pt}
\epsffile{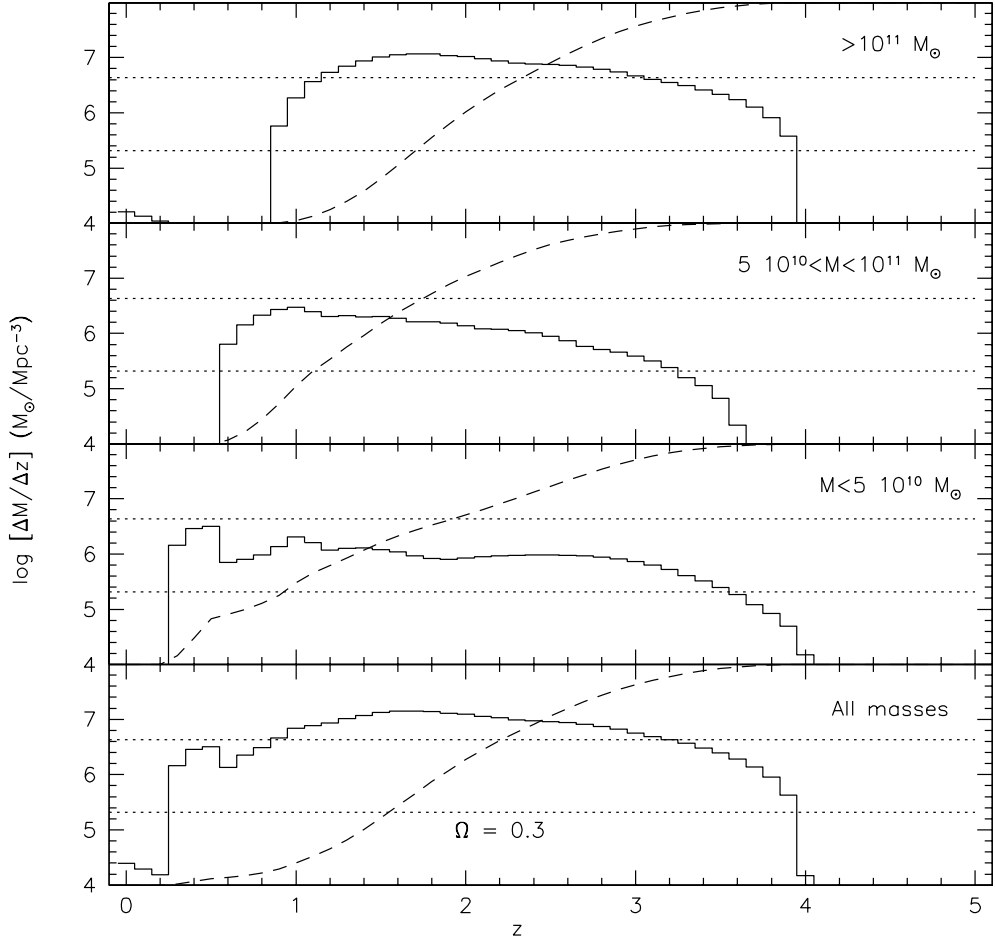}
\vspace*{-10pt}
\caption{Distributions of the density of stellar mass per unit comoving 
volume generated in the redshift bin as a function of redshift. 
The units are solar masses per cubic Mpc born in the redshift bin.
The distributions are splitted into the contributions    of various 
galactic mass ranges.
The dashed lines in the various panels report the cumulative distributions
of the mass density generated as a function of redshift, on a linear 
vertical scale ranging from 0\% to 100\%. 
The dotted horizontal lines mark the 33 and 66\% percentiles.
A value $q_0=0.15$ is adopted.
}
\label{fig16}
\end{figure}

\clearpage\newpage

\epsfxsize=14cm
\begin{figure}[!Ht]
\vspace*{-10pt}
\hspace*{0pt}
\epsffile{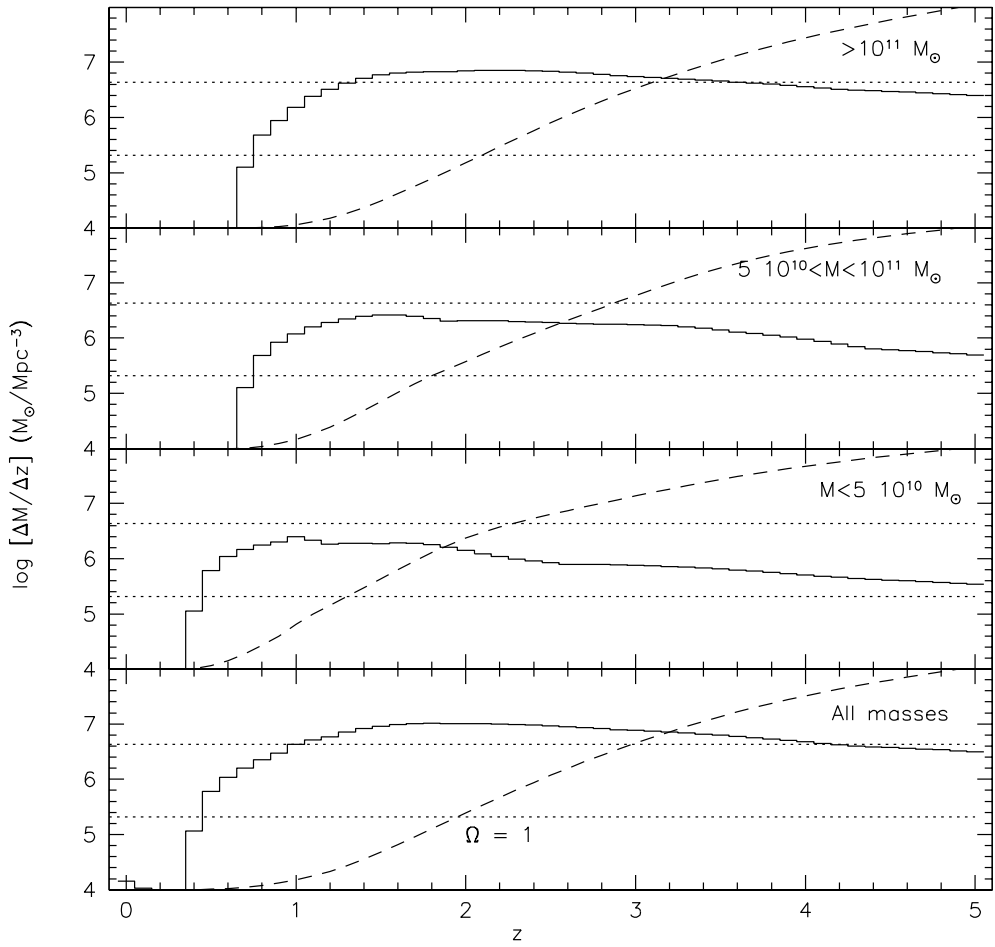}
\vspace*{-10pt}
\caption{Same as in Figure 16 but for $q_0=0.5$.
}
\label{fig17}
\end{figure}

\clearpage\newpage

\epsfxsize=14cm
\begin{figure}[!Ht]
\vspace*{-10pt}
\hspace*{0pt}
\epsffile{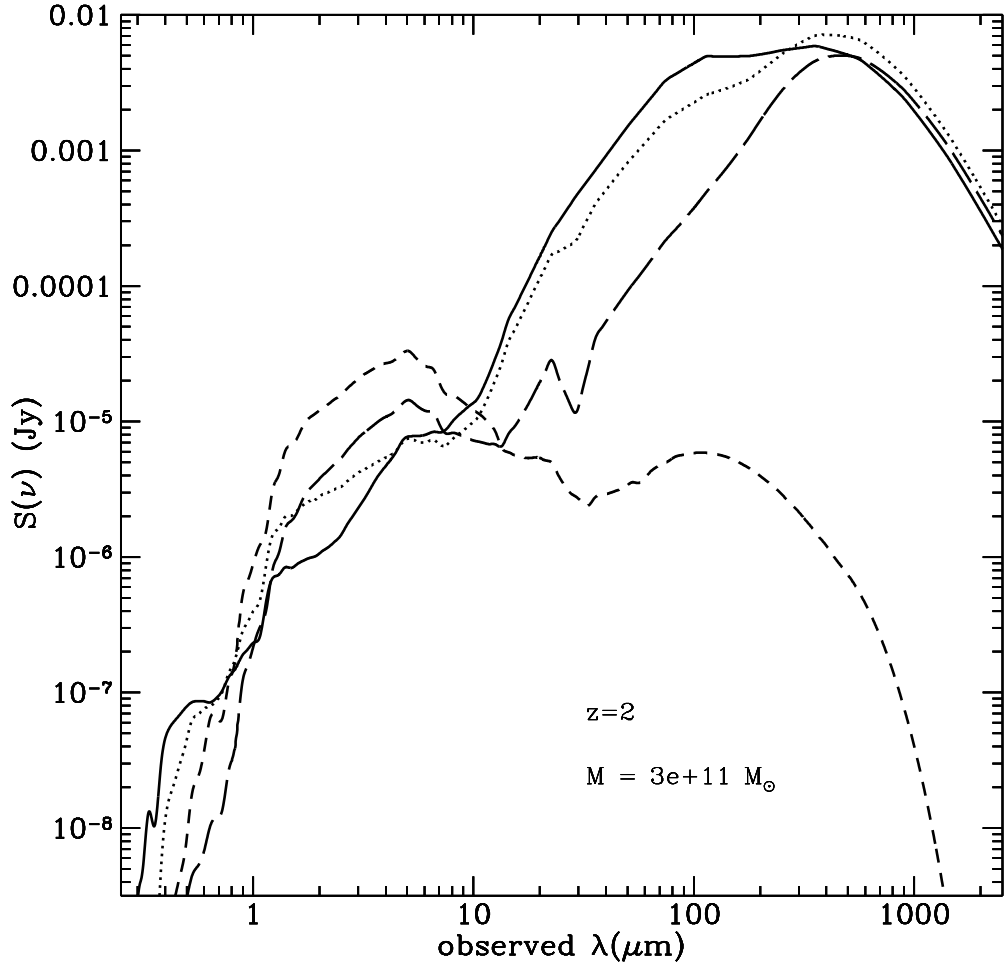}
\vspace*{-10pt}
\caption{Predicted spectra of a $M=3\ 10^{11}\ M_\odot$ galaxy at $z=2$
observed at 0.5 (continuous), 1 (dotted), 3 (long dash), and 4 (short 
dash) Gyr, according to Model 2. The former three are spectra during 
the dusty starburst phase.
}
\label{fig18}
\end{figure}
\clearpage\newpage

\epsfxsize=20cm
\begin{figure}[!Ht]
\vspace*{-10pt}
\hspace*{0pt}
\epsffile{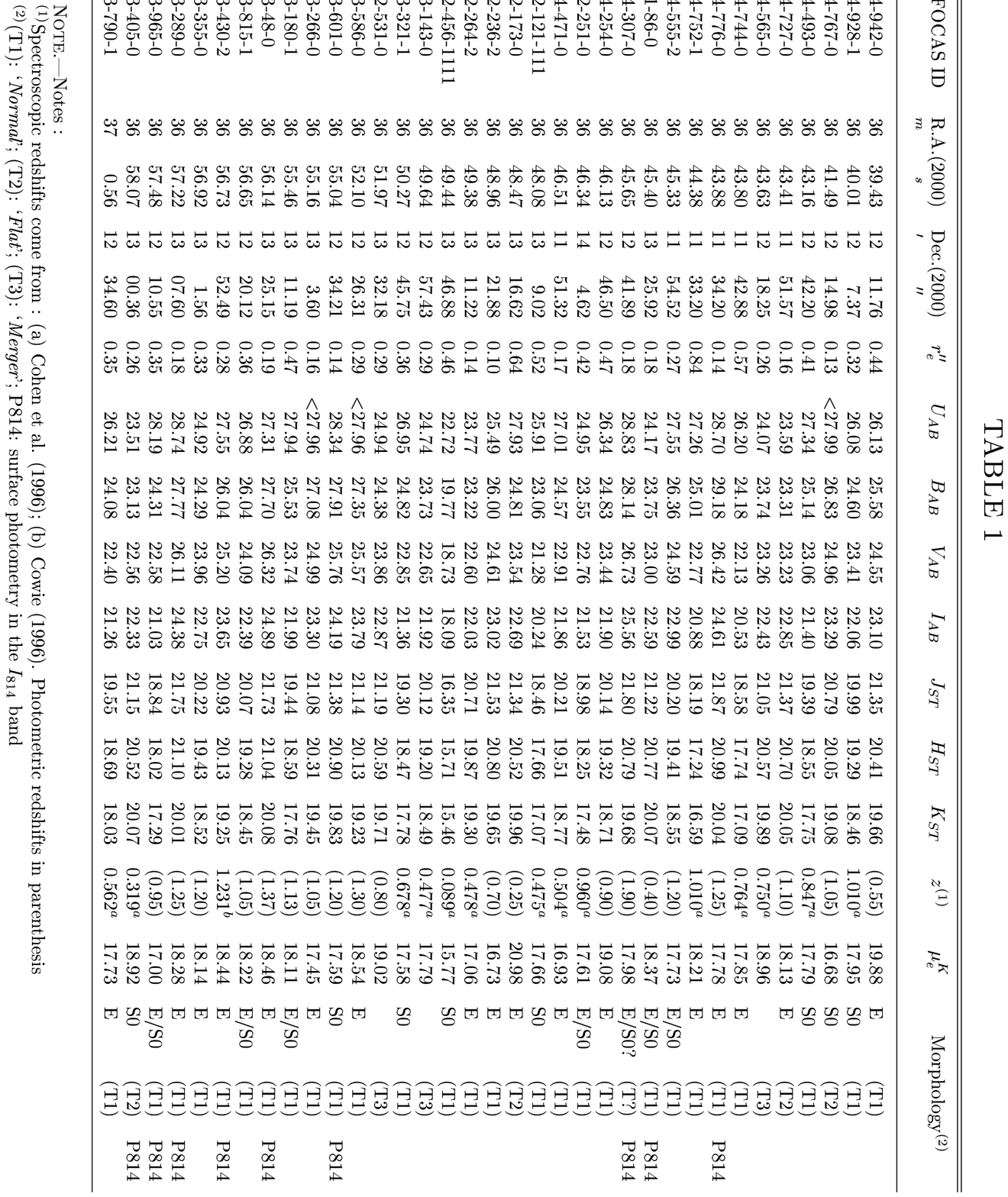}
\vspace*{-10pt}
\label{tab1}
\end{figure}

\clearpage\newpage

\epsfxsize=20cm
\begin{figure}[!Ht]
\vspace*{-10pt}
\hspace*{0pt}
\epsffile{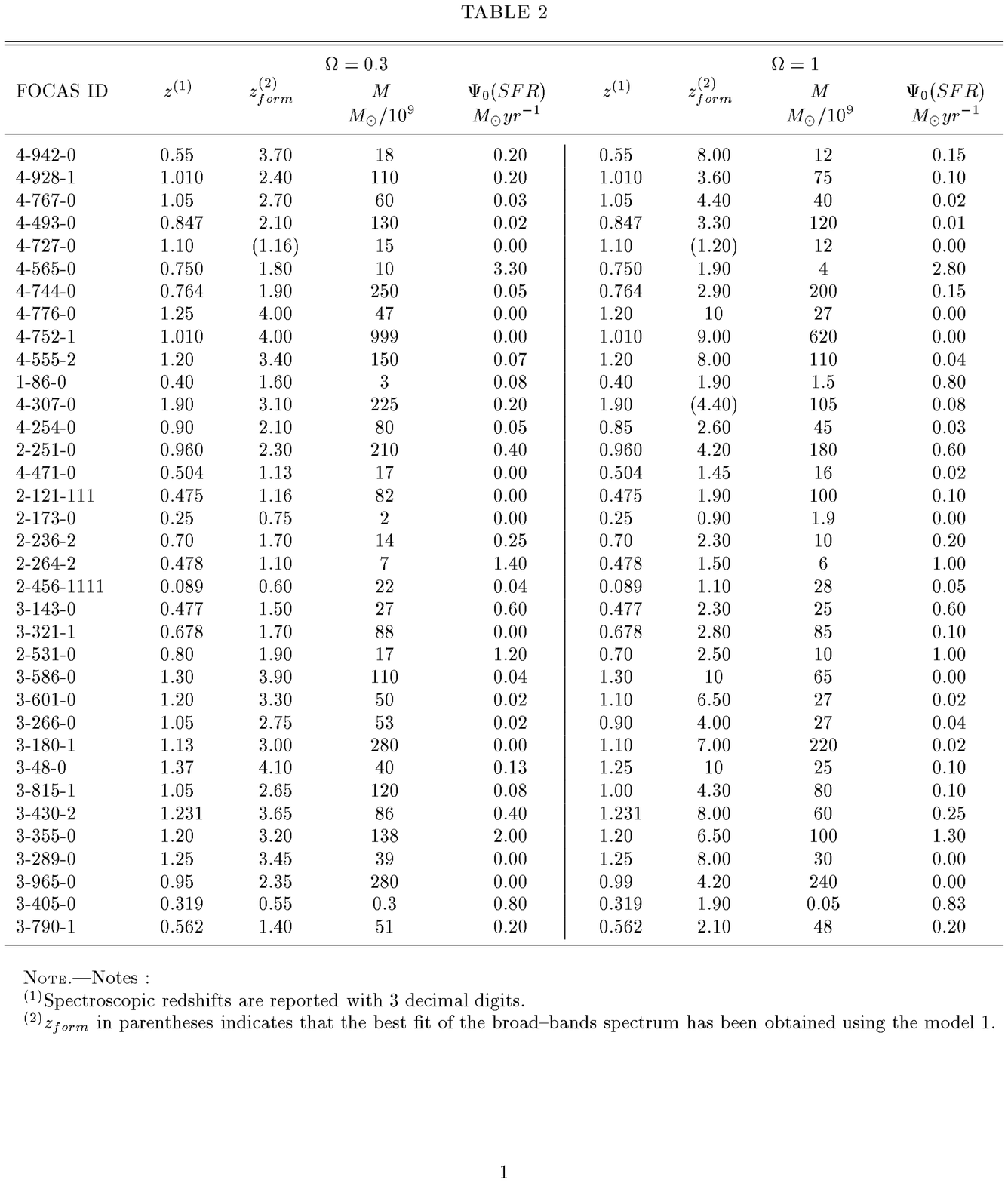}
\vspace*{-10pt}
\label{tab2}
\end{figure}

\end{document}